\RequirePackage{fix-cm}
\documentclass[twocolumn,epjc3]{svjour3}  
\smartqed  
\RequirePackage{graphicx}
\journalname{Eur. Phys. J. C}
\usepackage{cite}
\usepackage{amsmath}
\usepackage{amsfonts}
\usepackage{amssymb}
\usepackage{makeidx}
\usepackage{graphicx}
\usepackage{lmodern}
\usepackage{color}

\usepackage{bm}
\usepackage{enumerate}
\usepackage{float}
\usepackage{soul}
\usepackage{cancel}
\usepackage{multirow}
\usepackage{rotating}   
\usepackage{multirow} 
\usepackage{makecell}
\usepackage{booktabs}
\usepackage{bm}
\usepackage{subcaption} 
\usepackage{array}  

\begin{document}
\title{Metric Variation of the Energy-Momentum Tensor of a Perfect Fluid and Its Applications to Cosmology and Neutron Stars}

\author{Pham Van Ky \thanksref{e,addr1,addr2}}

\thankstext{e}{e-mail: phamkyvatly@gmail.com (corresponding author)}

\institute{
\small
\textit{Institute of physics},\\ 
\small
\textit{Vietnam academy of science and technology (VAST)},\\ 
\small
\textit{10 Dao Tan, Hanoi, Viet Nam.\\}\label{addr1}
\and
\small
\textit{Graduate university of science and technology},\\ 
\small
{\it Vietnam academy of science and technology (VAST),}\\ 
\small
{\it 18 Hoang Quoc Viet, Hanoi, Viet Nam.}\label{addr2}
}

\date{Received: date / Accepted: date}
%
%
\maketitle
\begin{abstract}
We show that the expressions for the matter Lagrangian \(L_m\) and the metric variation \(\delta T_{\mu\nu}\) of a perfect fluid obtained in previous studies appear to be inconsistent with the standard energy-momentum tensor under general conditions. Consequently, a large number of studies in astrophysics and cosmology relying on these expressions may need to be re-examined. By performing a series of straightforward calculations directly on the standard energy-momentum tensor \(T_{\mu\nu} = (\epsilon + P) u_\mu u_\nu - P g_{\mu\nu}\) together with the particle number conservation condition, we derive an expression for \(\delta T_{\mu\nu}\) that is independent of the choice of \(L_m\). Applying this result to \(f(R,T)\) gravity, we obtain the exact form of the tensor \(\Theta_{\mu\nu} = g^{\sigma\rho} \frac{\delta T_{\sigma\rho}}{\delta g^{\mu\nu}}\), which remains an important yet long-standing controversial quantity. This expression is shown to hold also for radiation, regardless of whether particle number is conserved. A major result is that if the energy-momentum tensor \(T_{\mu\nu}\) of the Universe consists solely of standard components with EOS \(P = \omega \epsilon\) where \(\omega = 0\), \(1/3\), or \(-1\) (baryonic/cold dark matter, radiation, and the cosmological constant), then \(f(R,T)\) gravity satisfies the conservation law \(\nabla_\mu T^{\mu\nu} = 0\) for any function \(f(R,T)\). This contrasts with previous studies, which found that the conservation law holds only for a restricted class of \(f(R,T)\) functions. Applying the same formalism to stellar interiors, we derive a class of functions that preserve the conservation law. We construct a specific \(f(R,T)\) model that is consistent at both cosmological scales and in high-density objects such as neutron stars. Remarkably, the same parameter set in this model simultaneously alleviates the Hubble tension and reproduces the observed mass-radius (M--R) relation of neutron stars.
\end{abstract}

\section{Introduction}
The Lagrangians for fields such as the electromagnetic field, scalar field, fermion field, etc., have been clearly established. However, there remains ambiguity regarding the appropriate form of the matter Lagrangian \(L_m\) that corresponds to the energy-momentum tensor of a perfect fluid, \(T_{\mu\nu} = (\epsilon + P) u_\mu u_\nu - P g_{\mu\nu}\) (the standard form). This ambiguity has caused considerable controversy and hindered the coherent development of the theory in both cosmology and astrophysics. 

In 2012, Minazzoli and Harko proposed the expression \(L_m = -\epsilon(\rho)\) with \(\epsilon(\rho) = \rho + \rho \int \frac{P(\rho)}{\rho^2} d\rho\) for a barotropic perfect fluid with particle number conservation \cite{Minazzoli:2012md}. Here, \(\rho\) is the rest-mass density, \(\epsilon(\rho)\) is the total rest-frame energy density, and \(P\) is the corresponding pressure (with \(c=1\)). However, this expression does not easily yield the metric variation of the energy-momentum tensor \(T_{\mu\nu}\) with respect to the metric tensor \(g_{\alpha\beta}\); this variation is denoted by \(\delta T_{\mu\nu}\). Subsequently, in 2024, Haghani, Harko, and Shahidi derived an expression for \(\delta T_{\mu\nu}\) by imposing thermodynamic conditions \cite{Haghani:2023uad}. 

Nevertheless, in most previous studies, including \cite{Minazzoli:2012md, Haghani:2023uad} and many others (e.g., \cite{Carvalho:2019gzs, Pretel:2021kgl, dosSantos:2018nmu, Bertini:2023pmp, Alvarenga:2013syu, Baffou:2013dpa}), it has been assumed that the matter Lagrangian \(L_m\) for a perfect baryonic fluid does not depend on the derivatives of the metric tensor, i.e., $\frac{\partial L_m}{\partial (\partial_\alpha g_{\mu\nu})} = 0.$ Consequently, the simplified expression \(T_{\mu\nu} = -g_{\mu\nu}L_m + 2\frac{\partial L_m}{\partial g^{\mu\nu}}\) has been widely used. In addition, some studies further impose the condition \(g^{\alpha\beta} \frac{\partial^2 L_m}{\partial g^{\mu\nu} \partial g^{\alpha\beta}} = 0\) \cite{Carvalho:2019gzs, Pretel:2021kgl, dosSantos:2018nmu, Bertini:2023pmp} and various thermodynamic conditions \cite{Haghani:2023uad}. 

Such conditions are valid for scalar fields and the electromagnetic field, but they are not appropriate for a perfect baryonic fluid. The reason is as follows: the matter Lagrangian \(L_m\) for a perfect baryonic fluid must at least include fields such as fermion fields, electromagnetic fields, and their interactions in a curved spacetime (gravitational field). That is, it contains terms involving the covariant derivatives of the fermion fields, \(\nabla_\mu \psi\). As is well known \cite{Birrell:1982ix}, the expression for \(\nabla_\mu \psi\) depends on the Christoffel symbols \(\Gamma^\alpha_{\mu\nu}\). Therefore, \(L_m\) must depend on the derivatives of the metric tensor. It should be emphasized that the expression for \(\rho\) itself implicitly contains dependence on both the metric tensor and its derivatives.

As will be shown in this article, imposing the aforementioned conditions on \(L_m\) leads to significantly different values for the relevant quantities compared to the case where no such conditions are imposed. This demonstrates that the matter Lagrangian \(L_m\) of a perfect baryonic fluid must depend substantially on the derivatives of the metric tensor.\\

In this work, we derive an expression for \(\delta T_{\mu\nu}\) of a perfect fluid with particle number conservation without imposing any conditions on the Lagrangian \(L_m\) or any thermodynamic constraints. The derivation is performed directly from the standard form of the energy-momentum tensor of a perfect fluid, using a series of simple calculations that rely only on well-established variational relations for the rest-mass density \(\rho\) and the four-velocity \(u^\mu\). The resulting expression for \(\delta T_{\mu\nu}\) is general and independent of the choice of the matter Lagrangian \(L_m\). It can be applied to modified gravity theories in which the gravitational Lagrangian contains the energy-momentum tensor \(T_{\mu\nu}\). By completely eliminating the dependence on \(L_m\), this work provides a clearer and more robust approach for studying modified theories of gravity in both cosmology and astrophysics.\\

To illustrate the usefulness of the derived \(\delta T_{\mu\nu}\), we apply it to obtain the tensor \(\Theta_{\mu\nu} = g^{\sigma\rho} \frac{\delta T_{\sigma\rho}}{\delta g^{\mu\nu}}\) in \(f(R,T)\) gravity, a quantity that remains controversial in the current literature. The \(f(R,T)\) gravity theory was proposed by Harko et al. in 2011 \cite{Harko:2011kv} as a natural extension of \(f(R)\) gravity \cite{DeFelice:2010aj, thomas, Ky:2025pax, Ky:2024lce, VanKy:2022itq, AnhKy:2018pbb} by allowing a coupling between the Ricci scalar \(R\) and the trace \(T = g^{\mu\nu} T_{\mu\nu}\) of the energy-momentum tensor of matter. The field equations of this theory contain an important additional tensor \(\Theta_{\mu\nu} = g^{\sigma\rho} \frac{\delta T_{\sigma\rho}}{\delta g^{\mu\nu}}\) arising from the variation of $ T_{\sigma\rho} $ with respect to the metric $ g^{\mu\nu} $. A major and persistent difficulty in \(f(R,T)\) gravity is the ambiguity in the choice of the matter Lagrangian \(L_m\). Different choices (\(L_m = P\) \cite{Harko:2011kv, Xu:2016rdf}, \(L_m = -\epsilon\) \cite{Bhattacharjee:2025tmn, Mota:2024kjb}, or more complicated forms) lead to different expressions for \(\Theta_{\mu\nu}\), even when using the same standard form of the perfect-fluid energy-momentum tensor \(T_{\mu\nu} = (\epsilon + P) u_\mu u_\nu - P g_{\mu\nu}\).

Recent attempts to construct \(f(R,T)\) gravity while avoiding the ambiguity in the choice of \(L_m\), such as in Ref.~\cite{Carvalho:2020czc}, start from a simplified condition in which \(T_{\mu\nu} + \Theta_{\mu\nu}\) is replaced by \(-T_{\mu\nu} - g_{\mu\nu} L_m\). This replacement is valid only if two conditions mentioned above are satisfied: first, \(L_m\) is independent of the derivatives of the metric tensor \(\frac{\partial L_m}{\partial (\partial_\alpha g_{\mu\nu})} = 0\), and second, \(g^{\alpha\beta} \frac{\partial^2 L_m}{\partial g^{\mu\nu} \partial g^{\alpha\beta}} = 0\). These imposed conditions limit the generality of the results obtained in that work. Notably, the authors concluded that the energy-momentum tensor of a perfect fluid cannot take the standard form.

In this article, we address this long-standing issue in \(f(R,T)\) gravity by applying our general expression for \(\delta T_{\mu\nu}\), from which the tensor \(\Theta_{\mu\nu}\) is obtained immediately. Most importantly, our results differ significantly from the values of \(\Theta_{\mu\nu}\) reported in previous studies, for example \cite{Haghani:2023uad, Carvalho:2019gzs, Pretel:2021kgl, dosSantos:2018nmu, Bertini:2023pmp}. This indicates that imposing the two conditions on the Lagrangian \(L_m\) leads to substantial deviations from the correct results. Therefore, previous studies need to be re-examined, both in cosmology and in astrophysics (including studies of stellar structure and compact stars such as neutron stars).

 To illustrate this re-examination, we apply the obtained \(\Theta_{\mu\nu}\) to investigate the interior structure of compact stars.
From this, we derive a specific form of the function \(f(R,T)\) that preserves the conservation law \(\nabla_\mu T^{\mu\nu} = 0\) for the single polytropic equation of state (EOS) \(P = K \epsilon^\Gamma\). We also obtain a modified Tolman--Oppenheimer--Volkoff (TOV) equation and compute the mass-radius (\(M\)--\(R\)) relation for neutron stars. From this, we derive a modified gravity model that can simultaneously reproduce the observed $M_{\rm max}$ and $R_{M_{\rm max}}$ of neutron stars.\\

In addition, by applying the obtained \(\Theta_{\mu\nu}\), we prove that the theory satisfies the conservation law \(\nabla_\mu T^{\mu\nu} = 0\) for any function \(f(R,T)\) when the matter inside the star obeys an equation of state with constant trace \(T = \text{constant}\) (for example, the special case \(P = \omega \epsilon\) with \(\omega = 1/3\), for which \(T = 0\)). This is a reasonable result because when \(T = \text{constant}\), the theory \(f(R,T)\) reduces to \(f(R)\) gravity, which is known to always satisfy \(\nabla_\mu T^{\mu\nu} = 0\). However, it can be easily verified that using the expressions for \(\Theta_{\mu\nu}\) from previous studies does not allow one to prove this property for arbitrary \(f(R,T)\); it holds only for one specific form of \(f(R,T)\). This is unreasonable and once again shows that imposing the aforementioned conditions on the Lagrangian \(L_m\) for a perfect baryonic fluid is not justified.\\

In General Relativity (GR), also known as the standard $ \Lambda CDM $ model, the Hubble parameter $H^{\Lambda CDM}(z)$ is expressed as
\begin{align}
\frac{H^{\Lambda CDM}(z)}{\sqrt{(1+z)^3\Omega^{\Lambda CDM}_{m, 0}+\Omega^{\Lambda CDM}_{\Lambda, 0}}}=H_0^{\Lambda CDM}. \label{GRct}
\end{align}
The superscript $ \Lambda CDM $ is introduced here to distinguish quantities in the standard $ \Lambda CDM $ model from those in modified gravity theories. The left-hand side of Eq.~\eqref{GRct} is a constant equal to $H^{\Lambda CDM}_0$. However, measurements of the Hubble constant using two independent methods reveal a significant discrepancy, commonly known as the Hubble tension: Early-universe methods (based on CMB data within the $\Lambda$CDM model) \cite{Planck:2018vyg, Hart:2019dxi} infer $H^{\Lambda CDM}(z\approx 1100)$ and, via Eq.~\eqref{GRct}, yield $H^{\Lambda CDM}_0 \approx 67.4 \, \mathrm{km\, s^{-1}\, Mpc^{-1}}$. Late-universe methods (cosmic distance ladder using Cepheids, Type Ia supernovae, and JWST) \cite{Riess:2021jrx, Breuval:2024lsv, Li:2025ife} give a direct local measurement at $z\approx 0$ of $H^{\Lambda CDM}_0 \approx 73$ $\mathrm{km\, s^{-1}\, Mpc^{-1}}$. This disagreement suggests that the $\Lambda$CDM model and standard gravity may be incomplete, motivating exploration of modified gravity as a potential resolution. In the standard $\Lambda$CDM model, $H_0$ (the quantity appearing on the right-hand side of Eq.~\eqref{GRct}) is referred to as the Hubble constant and is a fixed value. However, in modified gravity models proposed to resolve the Hubble tension, instead of a single constant value of $H_0$, one typically obtains a redshift-dependent quantity $\mathbb{H}_0(z)$. This quantity reflects the phenomenon of the running of $H_0$, where the value of $H_0$ inferred from observational data depends on the redshift $z$ at which it is measured. We denote this redshift-dependent quantity as $\mathbb{H}_0(z)$, following the notation of Ref.~\cite{Krishnan:2020vaf}.\\

We prove that the expression for \(\Theta_{\mu\nu}\) we obtained is valid not only for the case of particle number conservation but also for radiation (whether or not particle number is conserved). Applying this expression to cosmology, we obtain a key result: If the energy-momentum tensor \(T_{\mu\nu}\) of the Universe contains only standard matter components obeying equations of state (EOS) of the form \(P = \omega \epsilon\) with \(\omega = 0\), \(1/3\), or \(-1\) (e.g., baryonic matter and cold dark matter with \(\omega=0\), radiation with \(\omega=1/3\), and the cosmological constant as dark energy with \(\omega=-1\)), then \(f(R,T)\) gravity satisfies the conservation law \(\nabla_\mu T^{\mu\nu} = 0\) for any arbitrary function \(f(R,T)\). This is in contrast to previous studies, which concluded that the conservation law holds only for a restricted class of \(f(R,T)\) functions (e.g., \cite{Bertini:2023pmp, Alvarenga:2013syu, Baffou:2013dpa}). This result opens up a much richer space of viable \(f(R,T)\) functions for cosmological studies. In this paper, we apply the \(f(R,T)\) model derived from neutron star structure to cosmology. We find that the same parameter set simultaneously alleviates the Hubble tension while successfully reproducing the mass-radius (M--R) relation of neutron stars. Preliminary results also indicate that the model has the potential to reduce the \(S_8\) tension.\\

In this article the following conventions are used:

- Signature of the Minkowski metric: \((+,-,-,-)\), that is, the infinitesimal distance is given as 
  \[ ds^2 = \eta_{\mu\nu} dx^{\mu} dx^{\nu} \equiv (dx^0)^2 - dx^2 - dy^2 - dz^2,~ \text{with}~ x^0 = ct. \]

- Riemannian curvature tensor:
  \begin{align*}
  R^{\alpha}_{~\mu\beta\nu} = &\frac{\partial \Gamma^\alpha_{\mu\beta}}{\partial x^\nu} - \frac{\partial \Gamma^\alpha_{\mu\nu}}{\partial x^\beta} + \Gamma^\alpha_{\sigma\nu} \Gamma^\sigma_{\mu\beta} - \Gamma^\alpha_{\sigma\beta} \Gamma^\sigma_{\mu\nu}.
  \end{align*}

- Ricci tensor: \( R_{\mu\nu} = R^\alpha_{~\mu\alpha\nu} \).

- Scalar curvature: \( R = g^{\mu\nu} R_{\mu\nu} \), \( g^{\mu\nu} = g^{\mu\nu}(x) \).

- Energy-momentum tensor of the perfect fluid: 
  \begin{equation*}
  T_{\mu\nu} = \frac{1}{c^2} (\epsilon + P) u_\mu u_\nu - P g_{\mu\nu},
  \end{equation*}
  where \( u^\mu = \frac{dx^\mu}{d\tau} = c \frac{dx^\mu}{ds} \), \( u^\mu u_\mu = c^2 \), while \(\epsilon\) and \(P\) are the energy density and the pressure, respectively. 

- The Einstein equation
  \begin{align*}
  R_{\mu\nu} - \frac{1}{2} g_{\mu\nu} R = -k T_{\mu\nu},
  \end{align*}
  with \( k = 8\pi G / c^4 \).
  
- Note that the Lagrangian is more accurately referred to as the Lagrangian density. However, for the sake of brevity, we will continue to refer to it simply as the Lagrangian.

- Although the equation $\nabla_\mu T^{\mu\nu}=0$ does not strictly express the conservation of energy and momentum, it is conventionally called the conservation law.

\section{Metric Variation of the Energy-Momentum Tensor of a Perfect Fluid}

For a perfect fluid, the energy-momentum tensor of matter reads ($c=1$)
\begin{equation}
    T_{\mu\nu}= (\epsilon + P)u_\mu u_\nu - Pg_{\mu\nu}, \label{T1}
\end{equation}
with $u^\mu = \dfrac{dx^\mu}{d\tau} = \dfrac{dx^\mu}{ds}$. It is important to note that both $\epsilon$ and $P$ are quantities defined in the local rest frame of the fluid element. In particular, $\epsilon$ denotes the total rest-frame energy density, which consists of the rest-mass energy density associated with the rest mass density $\rho$ plus the internal energy density (which is related to the pressure). That is,
\begin{align}
& \epsilon = \epsilon(\rho), \label{mdn}\\
&P=P(\rho), \label{mda}\\
&\rho=\rho(x^\mu, g_{\alpha\beta}, \partial_\sigma g_{\alpha\beta}). \label{mdk}
\end{align}
Here, $\epsilon$ is the total rest energy density, $P$ is the pressure, and $\rho$ is the rest mass density. The dependence of $\epsilon$ and $P$ on the metric tensor and its derivatives is implicitly contained in the expression for the rest mass density $\rho$.

As discussed in the Introduction, there remains an ongoing debate regarding the most appropriate form of the matter Lagrangian $L_m$ that reproduces Eq.~\eqref{T1}. Therefore, in this work we do not derive the variation $\delta T_{\mu\nu}$ based on a specific choice of $L_m$. Instead, we obtain an expression for $\delta T_{\mu\nu}$ through an alternative route that is independent of the functional form of $L_m$.

The procedure is as follows, we rewrite Eq.~\eqref{T1} as
\begin{align}
T_{\mu\nu}= D_{\mu\nu} + E_{\mu\nu} - F_{\mu\nu}, \label{T2}
\end{align}
where
\begin{align}
&D_{\mu\nu}=\epsilon u_\mu u_\nu=\epsilon g_{\mu\alpha}g_{\nu\beta}u^\alpha u^\beta, \label{d2}\\
&E_{\mu\nu}=P u_\mu u_\nu=P g_{\mu\alpha}g_{\nu\beta}u^\alpha u^\beta, \label{d3}\\
&F_{\mu\nu}=Pg_{\mu\nu}. \label{d4}
\end{align}
The following relations hold:
\begin{align}
\delta D_{\mu\nu}=&g_{\mu\alpha}g_{\nu\beta}u^\alpha u^\beta\delta\epsilon + \epsilon g_{\nu\beta}u^\alpha u^\beta\delta g_{\mu\alpha}+\epsilon g_{\mu\alpha}u^\alpha u^\beta\delta g_{\nu\beta} \nonumber\\
&+\epsilon g_{\mu\alpha}g_{\nu\beta} u^\beta\delta u^\alpha + \epsilon g_{\mu\alpha}g_{\nu\beta} u^\alpha \delta u^\beta, \label{T3}
\end{align}
\begin{align}
&\delta E_{\mu\nu}=g_{\mu\alpha}g_{\nu\beta}u^\alpha u^\beta\delta P + P g_{\nu\beta}u^\alpha u^\beta\delta g_{\mu\alpha} \nonumber\\
&+P g_{\mu\alpha}u^\alpha u^\beta\delta g_{\nu\beta}+P g_{\mu\alpha}g_{\nu\beta} u^\beta\delta u^\alpha + P g_{\mu\alpha}g_{\nu\beta} u^\alpha \delta u^\beta, \label{p}
\end{align}
\begin{align}
\delta F_{\mu\nu}=g_{\mu\nu}\delta P + P\delta g_{\mu\nu}. \label{d5}
\end{align}
With Eqs.~\eqref{mdn}--\eqref{mdk}, we have
\begin{align}
&\delta P=\dfrac{\partial P}{\partial\rho}\delta\rho, \label{d6}\\
&\delta\epsilon=\dfrac{\partial\epsilon}{\partial\rho}\delta\rho. \label{d5a}
\end{align}
From the Appendix of Ref.~\cite{Harko:2010zi}, the variation of the rest mass density is 
\begin{align}
\delta \rho=\frac{1}{2}\rho(g_{\sigma\gamma}-u_\sigma u_\gamma)\delta g^{\sigma\gamma}. \label{T4}
\end{align}
It should be noted that Eq.~\eqref{T4} holds only under the assumption of particle number conservation, i.e., when $\nabla_\sigma (\rho u^\sigma) = 0$. This condition follows from the fact that the rest mass density $\rho$ is defined as the product of the particle number density and the rest mass per particle. Consequently, the relation $\nabla_\sigma (\rho u^\sigma) = 0$ represents the conservation of particle number.

Using the relation $g_{\mu\nu}g^{\nu\lambda}=\delta_\mu^\lambda$, one easily obtains
\begin{align}
\delta g_{\mu\alpha}=-g_{\mu\sigma}g_{\alpha\gamma}\delta g^{\sigma\gamma}. \label{T5}
\end{align}

Since the coordinate differentials $dx^\mu$ are primitive quantities, during the variation with respect to the metric components $g_{\sigma\gamma}$, we may treat $dx^\mu$ as independent of $g_{\sigma\gamma}$ (i.e., constant under metric variation). Note that $dx_\nu = g_{\nu\alpha} dx^\alpha$, so the lowered components $dx_\nu$ do depend on the metric. Employing $u^\mu = \dfrac{dx^\mu}{ds} = \dfrac{dx^\mu}{\sqrt{g_{\sigma\gamma}dx^\sigma dx^\gamma}}$ and $\delta u^\mu = \dfrac{\partial u^\mu}{\partial g_{\sigma\gamma}} \delta g_{\sigma\gamma}$, it follows that
\begin{align}
\delta u^\mu=-\frac{1}{2}u^\mu u^\sigma u^\gamma\delta g_{\sigma\gamma}=\frac{1}{2}u^\mu u_\sigma u_\gamma \delta g^{\sigma\gamma}. \label{T6}
\end{align}

Substituting Eqs.~\eqref{d6}--\eqref{T6} into Eqs.~\eqref{T3}--\eqref{d5}, we obtain the variation of the energy-momentum tensor of a perfect barotropic fluid as
\begin{align}
&\delta T_{\mu\nu}=\left\lbrace \dfrac{1}{2}\rho\left(g_{\sigma\gamma}-u_\sigma u_\gamma \right) \left[u_\mu u_\nu \dfrac{\partial \epsilon}{\partial\rho}+\left( u_\mu u_\nu-g_{\mu\nu}\right)\dfrac{\partial P}{\partial\rho}  \right] \right. \nonumber\\
&\left. \vphantom{\frac{X}{Y}} + \left(\epsilon +P \right) u_\gamma\left(u_\mu u_\nu u_\sigma -u_\mu g_{\nu\sigma}-u_\nu g_{\mu\sigma} \right) +Pg_{\gamma\nu}g_{\sigma\mu}\right\rbrace \delta g^{\sigma\gamma}, \label{bpt} 
\end{align}
or equivalently,
\begin{align}
\dfrac{\delta T_{\mu\nu}}{\delta g^{\sigma\gamma}}= \dfrac{1}{2}\rho\left(g_{\sigma\gamma}-u_\sigma u_\gamma \right) \left[u_\mu u_\nu \dfrac{\partial \epsilon}{\partial\rho}+\left( u_\mu u_\nu-g_{\mu\nu}\right)\dfrac{\partial P}{\partial\rho}  \right] \nonumber\\
 + \left(\epsilon +P \right) u_\gamma\left(u_\mu u_\nu u_\sigma -u_\mu g_{\nu\sigma}-u_\nu g_{\mu\sigma} \right) +Pg_{\gamma\nu}g_{\sigma\mu}. \label{bpt1} 
\end{align}

Equation \eqref{bpt} (and equivalently \eqref{bpt1}) represents the metric variation of the energy-momentum tensor of a perfect fluid with particle number conservation. It is general and independent of the matter Lagrangian $L_m$, and can therefore be used in any modified theory of gravity. In the future, if one constructs a modified gravity theory with non-minimal coupling between curvature tensors and $T_{\mu\nu}$ (for example, a Lagrangian containing a term such as $f(R, R^{\mu\nu}T_{\mu\nu})$ or similar terms involving $T_{\mu\nu}$), the variation of this Lagrangian with respect to the metric will benefit from Eq.~\eqref{bpt} to obtain the field equations without needing to specify $L_m$. This is useful because \[\delta f(R, R^{\mu\nu}T_{\mu\nu}) =\dfrac{\partial f}{\partial R}\delta R + \dfrac{\partial f}{\partial(R^{\mu\nu}T_{\mu\nu})}\delta(R^{\mu\nu}T_{\mu\nu}),\] where $\delta(R^{\mu\nu}T_{\mu\nu})=T_{\mu\nu}\delta R^{\mu\nu}+R^{\mu\nu}\delta T_{\mu\nu}$. Thus, one can directly use the expression for $\delta T_{\mu\nu}$ from Eq.~\eqref{bpt} in the field equations, without regard to the specific form of the matter Lagrangian $L_m$ that leads to the energy-momentum tensor $T_{\mu\nu}$.

Readers interested solely in the $\delta T_{\mu\nu}$ tensor may refer directly to this section without needing to read the entire manuscript. We hope that this general expression will be widely adopted by the scientific community in future studies of compact objects, cosmological dynamics, and other astrophysical phenomena.\\

To illustrate the usefulness of Eq.~\eqref{bpt1}, we now apply it to the definition $\Theta_{\sigma\gamma}=g^{\mu\nu}\dfrac{\delta T_{\mu\nu}}{\delta g^{\sigma\gamma}}$. One immediately obtains (noting that $u^\mu u_\mu=1$)
\begin{align}
\Theta_{\sigma\gamma}=\dfrac{1}{2}\rho\left( g_{\sigma\gamma}-u_\sigma u_\gamma\right)\dfrac{\partial(\epsilon-3P)}{\partial\rho}-T_{\sigma\gamma}. \label{T7}  
\end{align}
Note that, according to Eq.~\eqref{T1}, the trace is $T=\epsilon-3P$. Equation \eqref{T7} provides a completely general expression for the tensor $\Theta_{\sigma\gamma}$, which depends only on the standard form of the perfect-fluid energy-momentum tensor \eqref{T1} and is independent of the specific choice of the matter Lagrangian $L_m$. This expression can be directly applied to a wide range of problems in both cosmology and astrophysics. In cosmological applications, one may take $\dfrac{\partial P}{\partial\rho} = \dfrac{\dot{P}}{\dot{\rho}}$ and $\dfrac{\partial \epsilon}{\partial\rho} = \dfrac{\dot{\epsilon}}{\dot{\rho}}$; if $P=0$, then $\epsilon=\rho$.

If $u^1=u^2=u^3=0$ (as in the case of static gravitational sources such as stars and black holes, or in the comoving frame of the universe), the non-vanishing components of $\Theta^\mu_{~\nu}$ from Eq.~\eqref{T7} are (noting that $u^\mu u_\mu = u^0 u_0 = 1$)
\begin{align} 
&\Theta^0_{~0}=-T^0_{~0}=-\epsilon, \label{theta1}\\
&\Theta^1_{~1}=\Theta^2_{~2}=\Theta^3_{~3}=\frac{1}{2}\rho\dfrac{\partial(\epsilon-3P)}{\partial\rho} +P. \label{theta2}
\end{align}
These expressions hold both for static astrophysical objects (stars) and for the comoving universe. In the cosmological case with dust matter ($P=0$), one has $\epsilon=\rho$.

It is worth noting that, in general,
\begin{align}
T_{\mu\nu}&=\dfrac{2}{\sqrt{-g}}\dfrac{\delta(\sqrt{-g}L_m)}{\delta g^{\mu\nu}} \label{tl}\\
   &=-g_{\mu\nu}L_m+2\frac{\partial L_m}{\partial g^{\mu\nu}}-\frac{2}{\sqrt{-g}}\frac{\partial}{\partial{x^{\alpha}}}\left( \frac{\partial{(L_m\sqrt{-g})}}{\partial{\frac{\partial{g^{\mu\nu}}}{\partial{x^{\alpha}}}}}\right), \label{rt7}
\end{align}
\begin{align}
\Theta_{\mu\nu}&=g^{\sigma\rho}\frac{\delta T_{\sigma\rho}}{\delta g^{\mu\nu}}  \label{rt8}\\
&=-2T_{\mu\nu}-g_{\mu\nu}L_m+2g^{\alpha\beta}\frac{\partial^2L_m}{\partial g^{\mu\nu}\partial g^{\alpha\beta}}  \nonumber\\
&~~~~ -g^{\sigma\rho}\frac{\delta }{\delta g^{\mu\nu}}\left[\frac{2}{\sqrt{-g}}\frac{\partial}{\partial{x^{\alpha}}}\left( \frac{\partial{(L_m\sqrt{-g})}}{\partial{\frac{\partial{g^{\sigma\rho}}}{\partial{x^{\alpha}}}}}\right) \right]. \label{rt8d}
\end{align}
If one imposes the condition that $L_m$ is independent of the derivatives of the metric, then $T_{\mu\nu}= -g_{\mu\nu}L_m+2\frac{\partial L_m}{\partial g^{\mu\nu}}$, and the additional condition $g^{\alpha\beta}\frac{\partial^2L_m}{\partial g^{\mu\nu}\partial g^{\alpha\beta}} =0$, one obtains $\Theta_{\mu\nu}= -2T_{\mu\nu}-g_{\mu\nu}L_m$. Combining this with $L_m=-\epsilon$ (as discussed in the Introduction) yields $\Theta^1_{~1}=\epsilon + 2P$ \cite{Pretel:2021kgl, Carvalho:2019gzs, dosSantos:2018nmu}. This result differs significantly from the correct expression in Eq.~\eqref{theta2} (this difference is easily seen in the case $P=0$, where $\epsilon=\rho$).

Furthermore, the value in Eq.~\eqref{theta1} differs substantially from the result $\epsilon= -\dfrac{1}{2}(5\rho +3P)$ obtained in Ref.~\cite{Haghani:2023uad}. The expression in Eq.~\eqref{theta1} is accurate because it is derived without imposing any restrictive conditions on the Lagrangian $L_m$. In contrast, the result in Ref.~\cite{Haghani:2023uad} was obtained by imposing thermodynamic conditions and, in particular, the condition that $L_m$ does not depend on the derivatives of the metric tensor.\\

Thus, imposing the aforementioned conditions on the matter Lagrangian $L_m$ leads to results that differ significantly from the correct values. This demonstrates that, for a perfect baryonic fluid, one cannot neglect the dependence of $L_m$ on the derivatives of the metric tensor, i.e., $\dfrac{\partial g^{\mu\nu}}{\partial x^\alpha}$. Even in standard General Relativity (GR), the widely used simplified expression $T_{\mu\nu} = -g_{\mu\nu}L_m + 2\frac{\partial L_m}{\partial g^{\mu\nu}}$ may not be appropriate for a perfect baryonic fluid.\\

The expression \eqref{T7} has been derived under the assumption of particle number conservation. We now show that it also holds for radiation, whether or not particle number is conserved. Indeed, the Lagrangian of the electromagnetic field \(L^{(r)}\) is (restoring the speed of light \(c\)) \cite{Ky:2025pax, L2}
\begin{equation}
L^{(r)}=-\frac{1}{4}\varepsilon_0 c^2F^{\alpha\beta}F_{\alpha\beta}=-\frac{1}{4}\varepsilon_0 c^2g^{\alpha\sigma}g^{\beta\rho}F_{\sigma\rho}F_{\alpha\beta}, \label{T1e}
\end{equation}
from which we obtain the energy-momentum tensor of the electromagnetic field in a gravitational field:
\begin{equation}
T^{(r)}_{\mu\nu}=\varepsilon_0c^2\left(-F_{\mu\alpha}F_{\nu\beta}g^{\alpha\beta}+\frac{1}{4}F_{\alpha\beta}F^{\alpha\beta}g_{\mu\nu} \right). \label{TE}
\end{equation}
Here \( \frac{1}{4\pi\varepsilon_0}=9\times10^9~{\rm N\,m^2/C^2} \), and
\begin{equation}
F_{\mu\nu}=\nabla_\mu A_\nu-\nabla_\nu A_\mu=\partial_\mu A_\nu-\partial_\nu A_\mu, \label{F}
\end{equation}
where \(\nabla_\mu\) denotes the covariant derivative. From Eq.~\eqref{TE}, it is straightforward to see that
\begin{equation}
T^{(r)}=g^{\mu\nu}T^{(r)}_{\mu\nu}=T^{\mu(r)}_{\ \mu}=0. \label{TEa}
\end{equation}
Substituting \eqref{TE} into the definition \(\Theta_{\alpha\beta}=g^{\mu\nu}\dfrac{\delta T_{\mu\nu}}{\delta g^{\alpha\beta}}\), we readily obtain
\begin{align}
\Theta_{\alpha\beta}=-T_{\alpha\beta}. \label{Tr}
\end{align}
Equation \eqref{Tr} can also be recovered from Eq.~\eqref{T7}. Indeed, for radiation we have \(P=\epsilon/3\), which implies \(\epsilon-3P=0\) and \(\rho\dfrac{\partial(\epsilon-3P)}{\partial\rho}=0\). In this case, Eq.~\eqref{T7} reduces directly to Eq.~\eqref{Tr}. Thus, the expression \eqref{T7} holds for radiation as well.

When a physical system contains both types of matter, for example, baryonic matter (with particle number conservation) and radiation, with their respective energy-momentum tensors \(T^{(m)}_{\mu\nu}\) and \(T^{(r)}_{\mu\nu}\), the total quantities are \(T_{\mu\nu}= T^{(m)}_{\mu\nu}+ T^{(r)}_{\mu\nu}\) and \(\Theta_{\mu\nu}= \Theta^{(m)}_{\mu\nu}+ \Theta^{(r)}_{\mu\nu}\). Here, \(\Theta^{(m)}_{\mu\nu}\) takes the full form given by Eq.~\eqref{T7}, while \(\Theta^{(r)}_{\mu\nu}\) takes the simplified form \eqref{Tr}.

\section{Application to $f(R,T)$ Gravity \label{tfe}}

The action for $f(R,T)$ gravity is given by ($c=1$)
\begin{align}
S=\frac{1}{16\pi G}\int f(R, T)\sqrt{-g}\,d^4x + \int L_m \sqrt{-g}\,d^4x. \label{rt5}
\end{align}
Here, $R$ is the Ricci scalar, $T_{\mu\nu}$ is the energy-momentum tensor of matter, $T = g^{\mu\nu}T_{\mu\nu}$ is its trace, $f(R,T)$ is an arbitrary function of $R$ and $T$, and $L_m$ is the matter Lagrangian. Adopting the sign conventions stated in the Introduction, the variation of the action \eqref{rt5} with respect to the metric tensor yields the field equations (the sign conventions adopted in this article are identical to those in Ref.~\cite{Alvarenga:2013syu})
\begin{align}
&f_R(R, T)R_{\mu\nu}-\frac{1}{2}f(R, T)g_{\mu\nu}+\left( \nabla_\mu\nabla_\nu-g_{\mu\nu}\square \right) f_R(R, T)  \nonumber\\
&=-kT_{\mu\nu}-f_T(R, T)\left( T_{\mu\nu}+\Theta_{\mu\nu}\right), \label{rt6}
\end{align}
where $k=8\pi G$, $\nabla_\mu$ denotes the covariant derivative, $\square=\nabla_\mu\nabla^\mu$, $f_R(R, T)=\dfrac{\partial f(R, T)}{\partial R}$, $f_T(R, T)= \dfrac{\partial f(R, T)}{\partial T}$, $T_{\mu\nu}$ is given by Eqs.~\eqref{tl}--\eqref{rt7}, and $\Theta_{\mu\nu}$ is given by Eqs.~\eqref{rt8}--\eqref{rt8d}.

It is useful to recall the following identities:
\begin{align}
\left( \square\nabla_\nu-\nabla_\nu\square\right) \phi(R, T)=-R_{\nu\alpha}\nabla^\alpha\phi(R, T), \label{rt8a}
\end{align}
which holds for any scalar function $\phi(R, T)$, and the Bianchi identity
\begin{align}
\nabla^\mu\left( R_{\mu\nu}-\frac{1}{2}g_{\mu\nu}R\right) =0. \label{rt8b}
\end{align}

Applying the covariant divergence to the left-hand side of Eq.~\eqref{rt6} gives
\begin{align}
&\nabla^\mu\left[f_R(R, T)R_{\mu\nu}-\frac{1}{2}f(R, T)g_{\mu\nu} \right. \nonumber\\
&\left. \vphantom{\frac{X}{Y}} +\left( \nabla_\mu\nabla_\nu-g_{\mu\nu}\square \right) f_R(R, T) \right] = -\frac{1}{2}f_T(R, T)\nabla_\nu T. \label{rt8c}
\end{align}
In the original work by Harko et al.~\cite{Harko:2011kv}, the right-hand side of this equation is missing the term involving $\nabla_\nu T$.

Combining Eqs.~\eqref{rt6} and \eqref{rt8c} leads to
\begin{align}
&\nabla_\mu T^\mu_{~\nu}= \nonumber\\
&\frac{\left(T^\mu_{~\nu}+\Theta^\mu_{~\nu} \right)\nabla_\mu f_T(R, T)+ f_T( R, T)\left( \nabla_\mu\Theta^\mu_{~\nu}-\frac{1}{2}\nabla_\nu T\right) }{-k-f_T(R, T)}. \label{rt9cu}
\end{align}
This can be rewritten as
\begin{align}
&\nabla_\mu T^\mu_{~\nu}= \nonumber\\
&\frac{f_T( R, T)\left[ \left(T^\mu_{~\nu}+\Theta^\mu_{~\nu} \right)\nabla_\mu \ln f_T(R, T)+  \nabla_\mu\Theta^\mu_{~\nu}-\frac{1}{2}\nabla_\nu T \right] }{-k-f_T(R, T)}. \label{rt9}
\end{align}

\textbf{\subsection{Application to Stars \label{star}}}
For matter inside a static spherically symmetric star, according to Eqs.~\eqref{theta1}--\eqref{theta2}, the right-hand side of Eq.~\eqref{rt9} automatically vanishes for the components $\nu = (0, 2, 3)$. Therefore, in order to satisfy the conservation law $\nabla_\mu T^\mu_{~\nu}=0$, it is sufficient to require
\begin{align}
\left(T^\mu_{~1}+\Theta^\mu_{~1} \right)\nabla_\mu \ln f_T(R, T)+  \nabla_\mu\Theta^\mu_{~1}-\frac{1}{2}\nabla_\nu T=0 \label{grk}
\end{align}
(for $\nu=1$). Here we are looking for a model with $f(R,T)\neq f(R)$. According to Eqs.~\eqref{theta1}--\eqref{theta2}, substituting the explicit expressions for $\Theta^\mu_{~\nu}$, Eq. \eqref{grk} becomes
\begin{align}
&\frac{1}{2}\rho\dfrac{\partial(\epsilon-3P)}{\partial\rho} \partial_r \ln f_T(R, T)+  \partial_r\left[\frac{1}{2}\rho\dfrac{\partial(\epsilon-3P)}{\partial\rho} \right]\nonumber\\
& -\dfrac{1}{2}\partial_r\epsilon+\dfrac{5}{2}\partial_rP +\left( \Theta^1_{~1}-\Theta^0_{~0}\right)\Gamma^0_{01} =0. \label{ns1}
\end{align}
Care must be taken when deriving Eq.~\eqref{ns1}. For example, although $\Theta^0_{~1}=0$, one has $\nabla_0\Theta^0_{~1}=\left( \Theta^1_{~1}-\Theta^0_{~0}\right) \Gamma^0_{01}$. The requirement that the conservation law $\nabla_\mu T^{\mu\nu}=0$ holds for the matter is essential (see Appendix \ref{pla}).

For the static spherically symmetric metric
\begin{equation}
ds^2=e^{a(r)}{{dx^0}}^2-e^{b(r)}dr^2-r^2(d\theta^2+\sin^2\theta d\varphi^2), \label{dm4}
\end{equation}
where $x^0 = ct = t$, one also has the continuity equation for the radial component: $\nabla_\mu T^\mu_{~1} = -\partial_r P - (\epsilon + P)\dfrac{\partial_r a}{2}$. That is,
\begin{align}
-\partial_r P - (\epsilon + P)\dfrac{\partial_r a}{2} = 0. \label{ptlt}
\end{align}
Noting that $\Gamma^0_{01} = \dfrac{\partial_r a}{2}$, we substitute $\dfrac{\partial_r a}{2}$ from Eq.~\eqref{ptlt} into Eq.~\eqref{ns1} and use Eqs.~\eqref{theta1}--\eqref{theta2} to obtain
\begin{align}
&\rho\dfrac{\partial(\epsilon-3P)}{\partial\rho} \partial_r \ln f_T(R, T)+  \partial_r\left[\rho\dfrac{\partial(\epsilon-3P)}{\partial\rho} \right]\nonumber\\
& -\partial_r(\epsilon-3P) -\rho\dfrac{\partial(\epsilon-3P)}{\partial\rho}\dfrac{\partial_r P}{\epsilon+P} =0. \label{nsm}
\end{align}

We observe that if a star has an equation of state with constant trace $T = \epsilon - 3P=\rm constant$ (for example, the linear equation of state $P = \omega \epsilon$ with $\omega = 1/3$), then Eq.~\eqref{nsm} holds for any function $f(R,T)$. In other words, the conservation law $\nabla_\mu T^\mu_{~\nu}=0$ is satisfied for any $f(R,T)$ when $T=$ constant. This is reasonable because when $T=$ constant, the theory $f(R,T)$ effectively reduces to $f(R)$ gravity, which is known to preserve $\nabla_\mu T^{\mu\nu}=0$. This result confirms the correctness of the metric variation $\delta T_{\mu\nu}$ given in Eq.~\eqref{bpt} and the expression for $\Theta^\mu_{~\nu}$ in Eq.~\eqref{T7}. It is easy to check that if one uses the expressions for $\Theta^\mu_{~\nu}$ from previous studies (e.g., Refs.~\cite{Pretel:2021kgl, Carvalho:2019gzs, dosSantos:2018nmu, Bertini:2023pmp}), this property does not hold for arbitrary $f(R,T)$, but only for specific forms. This provides further evidence that the restrictive conditions previously imposed on the matter Lagrangian $L_m$ are not appropriate (see the discussion following Eq.~\eqref{rt8d}).\\

Now we consider models of the form
\begin{align}
f(R, T)=R+h(T), \label{fr1}
\end{align}
where $h(T)$ is an arbitrary function of $T$. Substituting into the field equations \eqref{rt6} and using the metric \eqref{dm4} together with Eqs.~\eqref{theta1}--\eqref{theta2}, we obtain
\begin{align}
-e^{-b(r)}\left( \frac{b'(r)}{r}-\frac{1}{r^2}\right) -\frac{1}{r^2}=-k\epsilon(r) + \frac{1}{2}h(T), \label{ns6}
\end{align}
\begin{align}
&e^{-b(r)}\left( \frac{a'(r)}{r}+\frac{1}{r^2}\right) -\frac{1}{r^2}= \nonumber\\
&~~~~~~~~~~~~~kP(r)+ \frac{1}{2}h(T)-\frac{1}{2}\rho\frac{\partial(\epsilon-3P)}{\partial \rho}h_T(T). \label{ns7}
\end{align}
Here $a'(r) = da(r)/dr$. Defining
\begin{align}
e^{-b(r)}=1-\frac{km(r)}{4\pi r}, \label{ns8}
\end{align}
Eq.~\eqref{ns6} becomes
\begin{align}
m'(r)=4\pi r^2\epsilon(r)-\frac{2\pi r^2}{k}h(T). \label{ns9}
\end{align}
Thus, the effective mass $m(r)$ inside a sphere of radius $r$ is
\begin{align}
m(r)= 4\pi\int_0^r r^2\epsilon(r)\,dr  -\dfrac{2\pi}{k}\int_0^r r^2 h(T)\,dr. \label{ns10cc}
\end{align}
The first term on the right-hand side is the standard GR contribution, while the second term represents the correction arising from the $f(R,T)$ theory.

Substituting Eqs.~\eqref{ptlt} and \eqref{ns8} into Eq.~\eqref{ns7}, we obtain the modified Tolman--Oppenheimer--Volkoff (TOV) equation:
\begin{align}
&P'(r)= -\dfrac{km(r)\epsilon(r)}{8\pi r^2}\left[1+\dfrac{P(r)}{\epsilon(r)} \right]\left[ 1-\dfrac{km(r)}{4\pi r}\right]^{-1}\times  \nonumber\\
& \left\lbrace 1+\dfrac{4\pi r^3P(r)}{m(r)}+\frac{2\pi r^3}{km(r)}\left[ h(T)-\rho\frac{\partial(\epsilon-3P)}{\partial\rho}h_T(T)\right] \right\rbrace . \label{tovcc}
\end{align}
Here $k=8\pi G$ and $m(r)$ is given by Eq.~\eqref{ns10cc}.\\

\subsubsection{Linear Equation of State $P=\omega \epsilon$}

We consider stars whose interior matter satisfies the linear equation of state $P = \omega \epsilon$ together with $\epsilon = \xi \rho$, where the constant $\xi$ depends on $\omega$, i.e., $\xi = \xi(\omega)$. In this case, Eq.~\eqref{nsm} reduces to
\begin{align}
(1-3\omega)\left[ \epsilon \partial_r \ln f_T(R, T) - \dfrac{\omega}{1+\omega}\partial_r \epsilon\right]  =0. \label{ns2}
\end{align}
We see that Eq.~\eqref{ns2} does not depend on the constant $\xi$, but only on the parameter $\omega$. This equation holds for any function $f(R, T)$ when $\omega = 1/3$. In this case, the trace of the energy-momentum tensor vanishes, $T=0$, which is a special case of the constant-trace condition $T=$ constant (as discussed below Eq.~\eqref{nsm}).

We now seek a model $f(R, T)$ that can describe arbitrary values of $\omega$. Setting the term in square brackets in Eq.~\eqref{ns2} to zero and solving the resulting differential equation, we obtain the general solution
\begin{align}
f_T(R, T)=\gamma\epsilon^{\frac{\omega}{1+\omega}}= \tilde{\gamma} T^{\frac{\omega}{1+\omega}}, \label{ns3}
\end{align}
where $\gamma$ is an arbitrary constant and $\tilde{\gamma}= \dfrac{\gamma}{(1-3\omega)^{\frac{\omega}{1+\omega}}}$. Integrating with respect to $T$ (noting that $T=\epsilon-3P=(1-3\omega)\epsilon$) yields
\begin{align}
f(R, T)=  X(R) + \tilde{\eta} T^{\frac{1+2\omega}{1+\omega}} = X(R) + \eta \epsilon^{\frac{1+2\omega}{1+\omega}}. \label{ns5}
\end{align}
Here the constants are $\tilde{\eta}=\frac{1+\omega}{1+2\omega}\tilde{\gamma}$ and $\eta=\frac{(1+\omega)(1-3\omega)}{1+2\omega}\gamma$, and \(X(R)\) is an arbitrary function of \(R\). Thus, we have obtained a general form of $f(R, T)$ given by Eq.~\eqref{ns5} that satisfies the conservation law $\nabla_\mu T^\mu_{~\nu}=0$ for stellar matter obeying the linear equation of state $P=\omega \epsilon$. This form differs significantly from those obtained in previous studies \cite{Carvalho:2019gzs, Pretel:2021kgl, dosSantos:2018nmu, Bertini:2023pmp}.

For simplicity, we now consider the simplest case $ X(R)=R$, i.e.,
\begin{align}
f(R, T)= R + \tilde{\eta} T^{\frac{1+2\omega}{1+\omega}} =R + \eta \epsilon^{\frac{1+2\omega}{1+\omega}}, \label{ns5a}
\end{align}
with $\tilde{\eta}= \frac{(1+\omega)\gamma}{(1+2\omega)(1-3\omega)^{\frac{\omega}{1+\omega}}}$ and $\eta=\frac{(1+\omega)(1-3\omega)\gamma}{1+2\omega}$. In this case, the effective mass function from Eq.~\eqref{ns10cc} becomes
\begin{align}
m(r)=& 4\pi\int_0^r r^2\epsilon(r)\,dr \nonumber\\
& -\dfrac{2\pi(1+\omega)(1-3\omega)\gamma}{k(1+2\omega)}\int_0^r r^2\epsilon^{\frac{1+2\omega}{1+\omega}}(r)\,dr. \label{ns10c}
\end{align}
Substituting into the modified TOV equation \eqref{tovcc}, we obtain
\begin{align}
&P'(r)= -\dfrac{km(r)\epsilon(r)}{8\pi r^2}\left[1+\dfrac{P(r)}{\epsilon(r)} \right]\left[ 1-\dfrac{km(r)}{4\pi r}\right]^{-1}\times  \nonumber\\
& \left[1+\dfrac{4\pi r^3P(r)}{m(r)}-\dfrac{2\pi \omega(1-3\omega)\gamma}{(1+3\omega)km(r)}r^3\epsilon^{\frac{1+2\omega}{1+\omega}}(r)\right]. \label{tovc}
\end{align}
Here $k=8\pi G$, $m(r)$ is given by Eq.~\eqref{ns10c}, and $P(r)=\omega\epsilon(r)$. Thus, Eq.~\eqref{tovc} is a differential equation for a single function $\epsilon(r)$.

For stars composed of ordinary (baryonic) matter, the parameter $\omega$ must satisfy $0 \leq \omega \leq 1$.

The numerical solution of Eq.~\eqref{tovc} proceeds as follows. First, a specific value of $\omega$ is chosen. Then, the central boundary condition $\epsilon(0) = \epsilon_c$ is imposed, where $\epsilon_c$ is the central total rest energy density. Solving the differential equation yields the radial profile $\epsilon(r)$. The stellar radius $R$ is determined by the condition $\epsilon(R) = 0$ (which implies $P(R) = 0$). Substituting the obtained $\epsilon(r)$ into Eq.~\eqref{ns10c} (with the upper limit $r = R$) gives the total mass $M$ of the star. Thus, each central density $\epsilon_c$ corresponds to a unique pair $(M, R)$. By varying $\epsilon_c$ over a wide range, one obtains the mass-radius ($M$--$R$) curve. The branch with $dM/dR > 0$ corresponds to stable configurations, while the branch with $dM/dR < 0$ is unstable. The maximum point on the curve, $(M_{\rm max}, R_{M_{\rm max}})$, defines the maximum mass $M_{\rm max}$ that a star with the chosen equation-of-state parameter $\omega$ can achieve. Any star with $M > M_{\rm max}$ will collapse into a black hole.

Note that we work in geometric units where $G = c = 1$. If the central energy density $\epsilon_c$ is given in SI units (J/m$^3$), it must be converted to geometric units via $\epsilon_{c\, \rm Geo} = 10^6 \times (G/c^4) \times \epsilon_{c\, \rm SI}$, yielding units of km$^{-2}$. In these units, lengths, times, and masses are all measured in km (since $[M] = [r] = [t] = [\epsilon]^{-1/2}$). The radius $R$ obtained directly from the numerical solution is already in km and corresponds to the physical radius in SI units. The mass $M$ (in km) is converted to solar masses by dividing by $M_\odot = 1.4766\, \rm km$.

After performing the numerical integration, we find that for any value of the parameter $\gamma$, Eqs.~\eqref{tovc}--\eqref{ns10c} yield no solution in which $\epsilon(r)$ (and thus $P(r)$) reaches zero at a finite radius $r$. This means that no value of $\gamma$ allows the linear equation of state $P=\omega\epsilon$ to describe a realistic star with a well-defined surface. The same conclusion holds in GR (recovered when $\gamma=0$).

\subsubsection{Polytropic Equation of State $P=K\epsilon^\Gamma$}

We consider stars whose interior matter satisfies the polytropic equation of state, which is conventionally written as 
\begin{align}
P=K\epsilon^\Gamma, \label{po1}
\end{align}
where $\Gamma \neq 1$ and $K$ are constants. As previously emphasized, the dependence of the matter Lagrangian $L_m$ on the derivatives of the metric tensor cannot be neglected. The precise form of this dependence depends on the types of matter and their interactions inside the star (note that $\rho$ itself depends on both the metric tensor and its derivatives, as shown in Eq.~\eqref{mdk}). Therefore, we must introduce an additional equation of state. From Eq.~\eqref{po1}, it is natural to assume that the stellar matter also satisfies
\begin{align}
\epsilon= A\rho^B, \label{po2}
\end{align}
where the constants $A$ and $B$ depend on $\Gamma$ and $K$, i.e., $A=A(\Gamma, K)$ and $B=B(\Gamma, K)$. From Eqs.~\eqref{po1} and \eqref{po2}, we have $P = K A^\Gamma \rho^{B\Gamma}$. Using these relations, we obtain
\begin{align}
\rho\frac{\partial(\epsilon-3P)}{\partial\rho}=(\epsilon-3K\Gamma \epsilon^\Gamma)B. \label{pop}
\end{align}
Substituting into Eq.~\eqref{nsm} yields
\begin{align}
&\partial_r\ln f_T(R, T)=\nonumber\\
&\left[\frac{k\Gamma\epsilon^{\Gamma-1}}{\epsilon+K\epsilon^\Gamma} -\frac{B-1-(B\Gamma-1)3K\Gamma\epsilon^{\Gamma-1}}{B(\epsilon-3K\Gamma\epsilon^\Gamma)}\right] \partial_r\epsilon. \label{po3} 
\end{align}

We see that Eq.~\eqref{po3} is independent of the constant $A$ in Eq.~\eqref{po2}. For convenience in integration, it can be rewritten as
\begin{align}
&\partial_r\ln f_T(R, T)=\left[ \frac{k\Gamma \epsilon^{\Gamma-2}}{1+K\epsilon^{\Gamma-1}}\right.\nonumber\\
&\left.-\frac{(B-1)(1-3K\Gamma^2\epsilon^{\Gamma-1})}{B(\epsilon-3K\Gamma\epsilon^\Gamma)}+\frac{3K\Gamma(\Gamma-1)\epsilon^{\Gamma-2}}{B(1-3K\Gamma\epsilon^{\Gamma-1})}\right] \partial_r\epsilon. \label{po4}
\end{align}
The general solution of Eq.~\eqref{po4} is
\begin{align}
f_T(R, T)=&\lambda\left( 1+K\epsilon^{\Gamma-1}\right)^{\frac{\Gamma}{\Gamma-1}}\left(\epsilon-3K\Gamma\epsilon^\Gamma \right)^{-\frac{B-1}{B}}\times \nonumber\\
&\left( 1-3K\Gamma\epsilon^{\Gamma-1}\right)^{\frac{-1}{B}}, \label{po5}
\end{align}
where $\lambda$ is an arbitrary constant. This can also be expressed as
\begin{align}
f_T(R, T)=\frac{\lambda\left( 1+K\epsilon^{\Gamma-1}\right)^{\frac{\Gamma}{\Gamma-1}}}{(1-3K\Gamma\epsilon^{\Gamma-1})\epsilon^{\frac{B-1}{B}}}. \label{po6}
\end{align}
Where $\epsilon$ is related to the trace $T$ by $T=\epsilon-3P$, we have
\begin{align}
T=\epsilon-3K\epsilon^\Gamma. \label{po7}
\end{align}
Using $f_T(R, T)=\dfrac{\partial f}{\partial T}=\dfrac{\partial f}{\partial\epsilon}\left( \dfrac{\partial T}{\partial\epsilon}\right)^{-1}$, Eq.~\eqref{po6} leads to
\begin{align}
\frac{\partial f(R, T)}{\partial\epsilon}=\lambda\left( 1+K\epsilon^{\Gamma-1}\right)^{\frac{\Gamma}{\Gamma-1}}\epsilon^{-\frac{B-1}{B}}. \label{po8}
\end{align}

To illustrate, we consider the simplest case with $B=1$ and $\Gamma=2$. Then Eq.~\eqref{po8} becomes
\begin{align}
f(R, T)= Y(R)+\frac{\lambda}{3K}\left( 1+K\epsilon\right)^3, \label{po9}
\end{align}
where $ Y(R)$ is an arbitrary function of $R$. In this case, Eq.~\eqref{po6} takes the form
\begin{align}
f_T(R, T)=\frac{\lambda\left( 1+K\epsilon\right)^2}{1-6K\epsilon}. \label{po10}
\end{align}
We observe that when the expression \(\frac{\lambda}{3K}\left(1 + K\epsilon\right)^3\) in model \eqref{po9} is expanded explicitly, it produces a minimum constant (uncoupled) term \(\frac{\lambda}{3K}\). However, in order to produce a significant modification to the stellar structure compared to General Relativity, the value of \(\frac{\lambda}{3K}\) must be quite large compared to the cosmological constant \(\Lambda\). Therefore, to obtain a more unified model that can provide substantial corrections to stellar structure relative to GR while remaining suitable for describing the evolution of the Universe and still satisfying Eq.~\eqref{po9}, we consider the following simple form:
\begin{align}
f(R, T) &= R - 2\Lambda - \frac{\lambda}{3K} + \frac{\lambda}{3K}\left(1 + K\epsilon\right)^3, 
\end{align}
which is equivalent to
\begin{align}
h(T) &= -2\Lambda - \frac{\lambda}{3K} + \frac{\lambda}{3K}\left(1 + K\epsilon\right)^3, \nonumber\\
     &= -2\Lambda + \lambda \epsilon\left(1 + K\epsilon + \frac{1}{3}K^2\epsilon^2 \right), \label{po11}
\end{align}
and
\begin{align}
h_T(T) = \frac{\lambda\left(1 + K\epsilon\right)^2}{1 - 6K\epsilon}. \label{po12}
\end{align}
Here \(\Lambda\) is the cosmological constant, and the relation between \(\epsilon\) and \(T\) according to \eqref{po7} is \(T = \epsilon - 3K\epsilon^2\). Obviously, this model satisfies the two important stability conditions \(\partial f(R,T)/\partial R \geq 0\) and \(\partial^2 f(R,T)/\partial R^2 \geq 0\) \cite{Dolgov:2003px, Faraoni:2006sy}.

The expression \eqref{po11} can be interpreted as follows: it is a model of the form \(f(R,T) = R + h(T)\) with
\begin{align}
h(T) = -2\Lambda + \lambda Q\left(1 + KQ + \frac{1}{3}K^2 Q^2 \right), \label{hn1}
\end{align}
where \(Q\) is a function of the trace \(T\) defined by
\begin{align}
T = Q - 3KQ^2. \label{hn2}
\end{align}
When this model is applied to describe stellar matter with the equation of state \(P = K\epsilon^2\), which implies \(T=\epsilon-3P = \epsilon - 3K\epsilon^2\), comparing with Eq.~\eqref{hn2} yields \(Q = \epsilon\). In this case, the model satisfies the conservation law \(\nabla_\mu T^\mu_{~\nu} = 0\). Note that this model satisfies \(Q = \epsilon\) only when the stellar matter obeys the equation of state \(P = K\epsilon^2\); under this condition, the conservation law \(\nabla_\mu T^\mu_{~\nu} = 0\) holds.\\

\begin{table*}[htbp]
\centering
\caption{\textit{Maximum mass and radius of neutron stars, along with related quantities, for different values of the parameter $\lambda$ with four representative polytropic constants $K=220$, $230$, $240$ and $250$. The table is presented in geometric units ($G=c=1$). Percentage changes are calculated relative to GR ($\lambda=0$).}}
\label{tab:3in1}

\begin{tabular}{l|cccccccc}
\hline
$K$ [km$^2$] & $\lambda$ & $M_{\rm max}$ [$M_\odot$] & $R_{\rm Mmax}$ [km] & $\epsilon_c$ [km$^{-2}$] at $ M_{\rm max} $ & $\Delta M_{\rm max}$ (\%) & $\Delta R_{\rm Mmax}$ (\%) & $T_c$ [km$^{-2}$] & $v_s^2=2K\epsilon_c$ \\
\hline
\multirow{4}{*}{220} 
& 0~(GR) & 2.004 & 11.17 & 0.001912 & +0.00 & +0.00 & -0.000501 & 0.8413 \\
& 6.218 & 2.198 & 11.83 & 0.002082 & +9.68 & +5.91 & -0.000779 & 0.9160 \\
& 7.416 & 2.243 & 11.96 & 0.002134 & +11.93 & +7.07 & -0.000871 & 0.9389 \\
& 8.566 & 2.289 & 12.09 & 0.002192 & +14.22 & +8.24 & -0.000979 & 0.9645 \\
\hline
\multirow{4}{*}{230} 
& 0~(GR) & 2.049 & 11.42 & 0.001828 & +0.00 & +0.00 & -0.000478 & 0.8409 \\
& 6.218 & 2.247 & 12.10 & 0.001991 & +9.66 & +5.95 & -0.000744 & 0.9157 \\
& 7.416 & 2.293 & 12.23 & 0.002040 & +11.91 & +7.09 & -0.000832 & 0.9386 \\
& 8.566 & 2.341 & 12.36 & 0.002096 & +14.25 & +8.23 & -0.000935 & 0.9641 \\
\hline
\multirow{4}{*}{240} 
& 0~(GR) & 2.093 & 11.67 & 0.001752 & +0.00 & +0.00 & -0.000458 & 0.8409 \\
& 6.218 & 2.296 & 12.36 & 0.001908 & +9.70 & +5.91 & -0.000713 & 0.9157 \\
& 7.416 & 2.343 & 12.49 & 0.001955 & +11.94 & +7.03 & -0.000797 & 0.9385 \\
& 8.566 & 2.391 & 12.62 & 0.002009 & +14.24 & +8.14 & -0.000896 & 0.9641 \\
\hline
\multirow{4}{*}{250} 
& 0~(GR) & 2.136 & 11.91 & 0.001683 & +0.00 & +0.00 & -0.000441 & 0.8413 \\
& 6.218 & 2.343 & 12.61 & 0.001832 & +9.69 & +5.88 & -0.000686 & 0.9161 \\
& 7.416 & 2.391 & 12.75 & 0.001878 & +11.94 & +7.05 & -0.000767 & 0.9390 \\
& 8.566 & 2.440 & 12.88 & 0.001929 & +14.23 & +8.14 & -0.000862 & 0.9646 \\
\hline
\end{tabular}
\end{table*}

\begin{figure*}[htbp]
\caption{\textit{The figures show the mass-radius ($M$--$R$) curves and the mass-central energy density ($M$--$\epsilon_c$) curves. The filled circles on each curve indicate the location of the maximum mass ($M_{\rm max}$). In each panel, moving from left to right along a curve, the mass first increases (stable branch) up to the maximum point and then decreases (unstable branch), corresponding to configurations that are gravitationally unstable and will collapse into black holes.}}
\label{fig:6in1}
\centering

\begin{subfigure}[b]{0.48\textwidth}
    \centering
    \includegraphics[width=\textwidth]{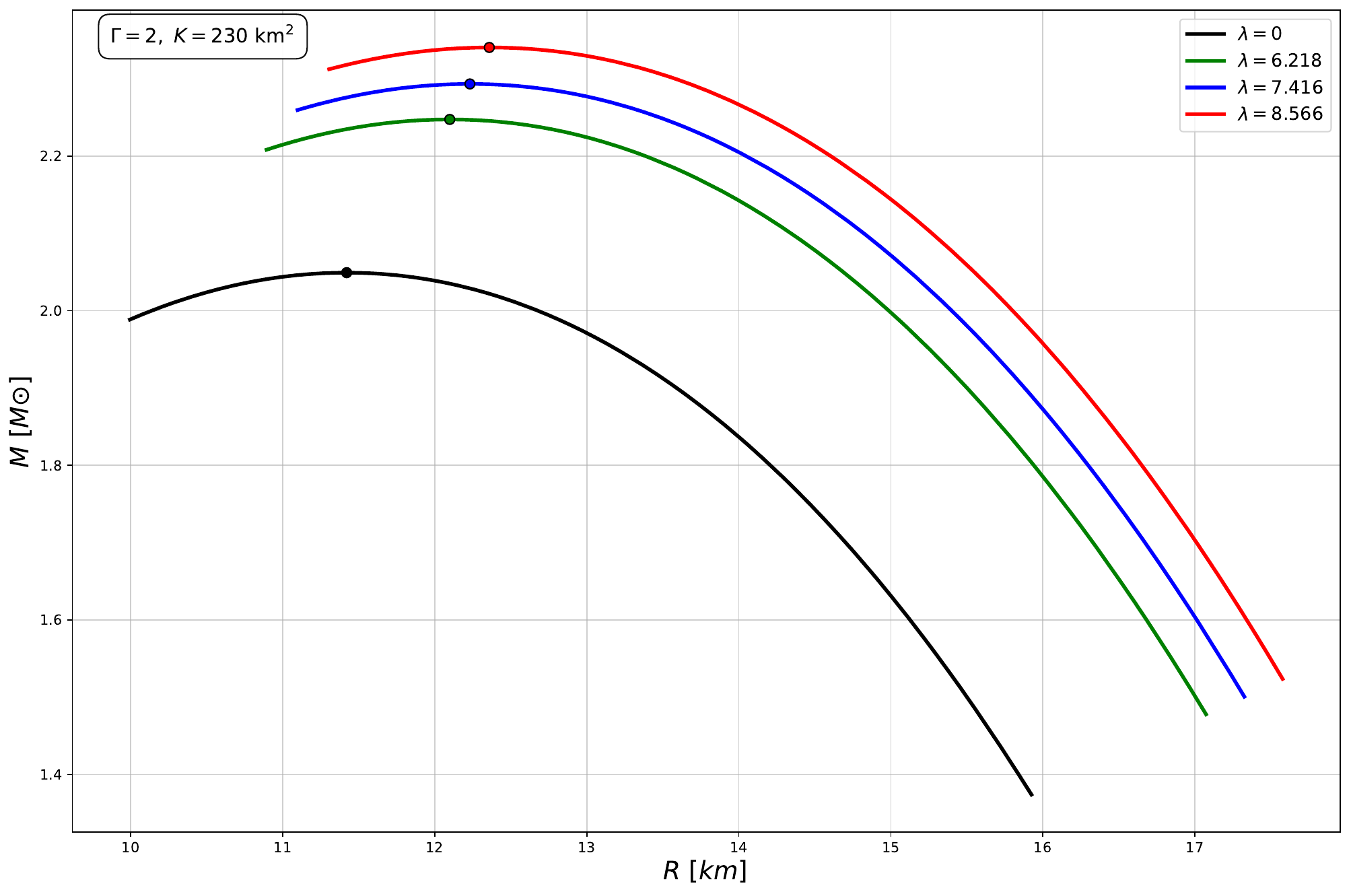}
\end{subfigure}
\hfill
\begin{subfigure}[b]{0.48\textwidth}
    \centering
    \includegraphics[width=\textwidth]{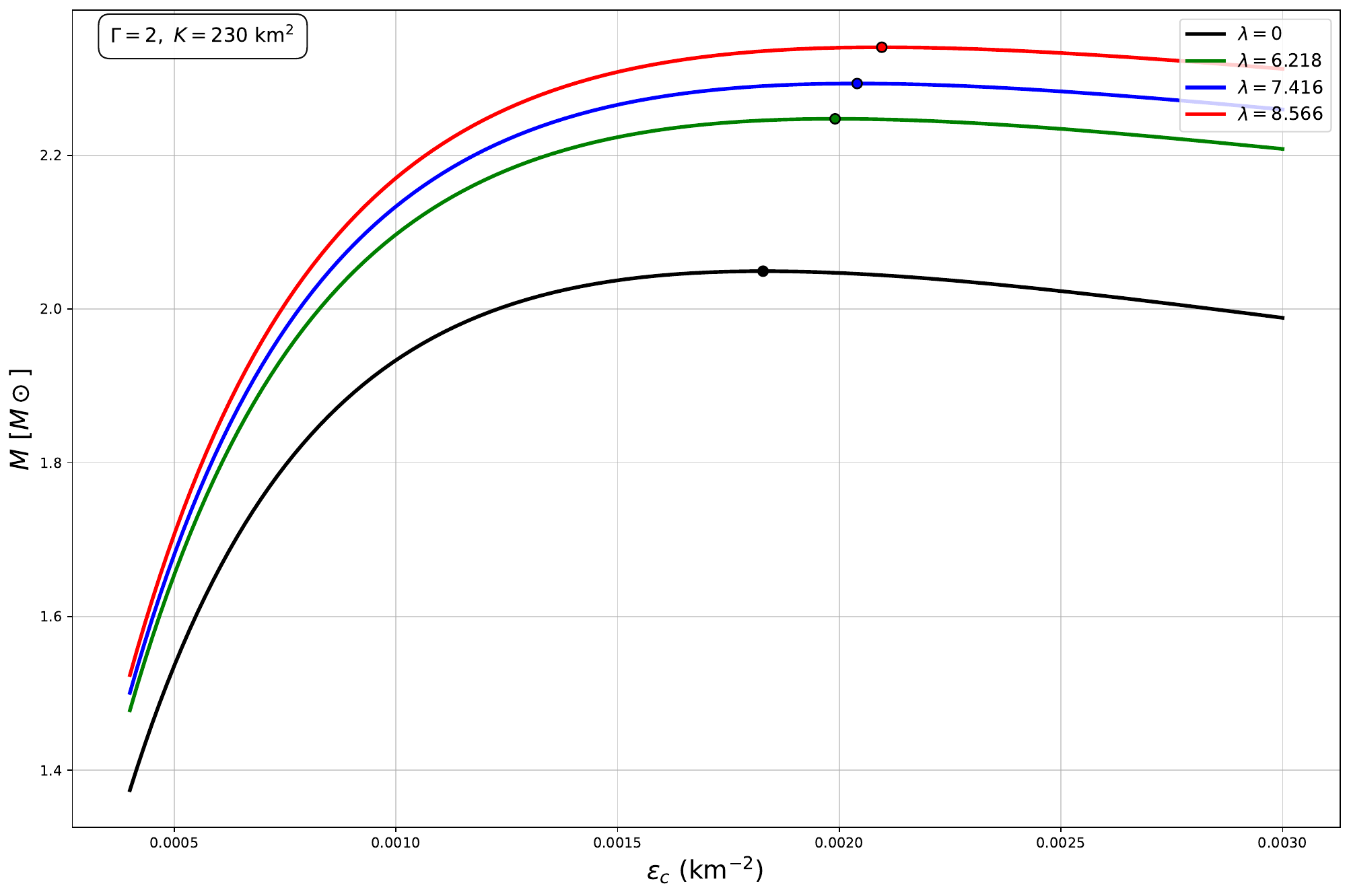}
\end{subfigure}

\vspace{1.2em} 

\begin{subfigure}[b]{0.48\textwidth}
    \centering
    \includegraphics[width=\textwidth]{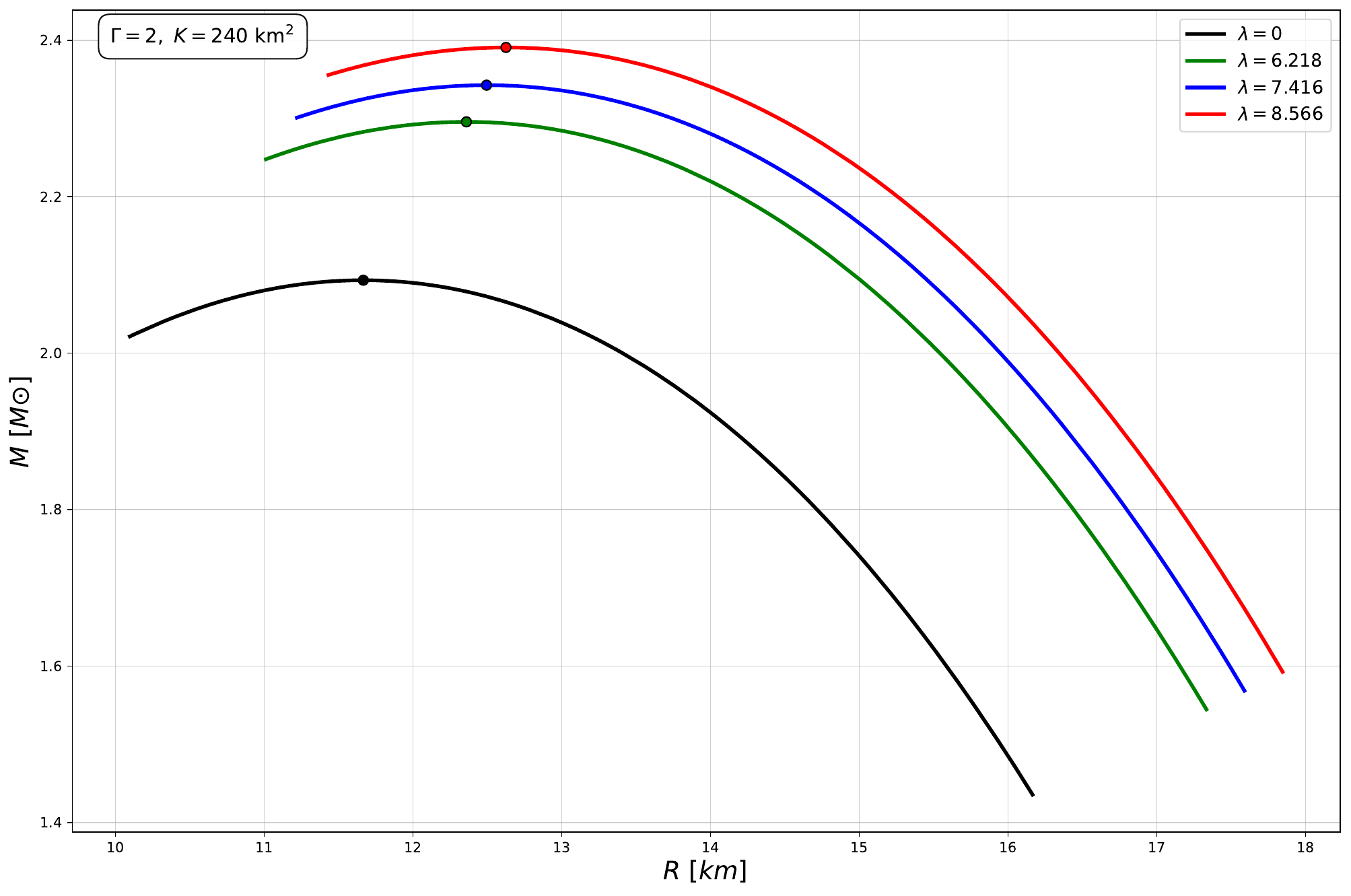}
\end{subfigure}
\hfill
\begin{subfigure}[b]{0.48\textwidth}
    \centering
    \includegraphics[width=\textwidth]{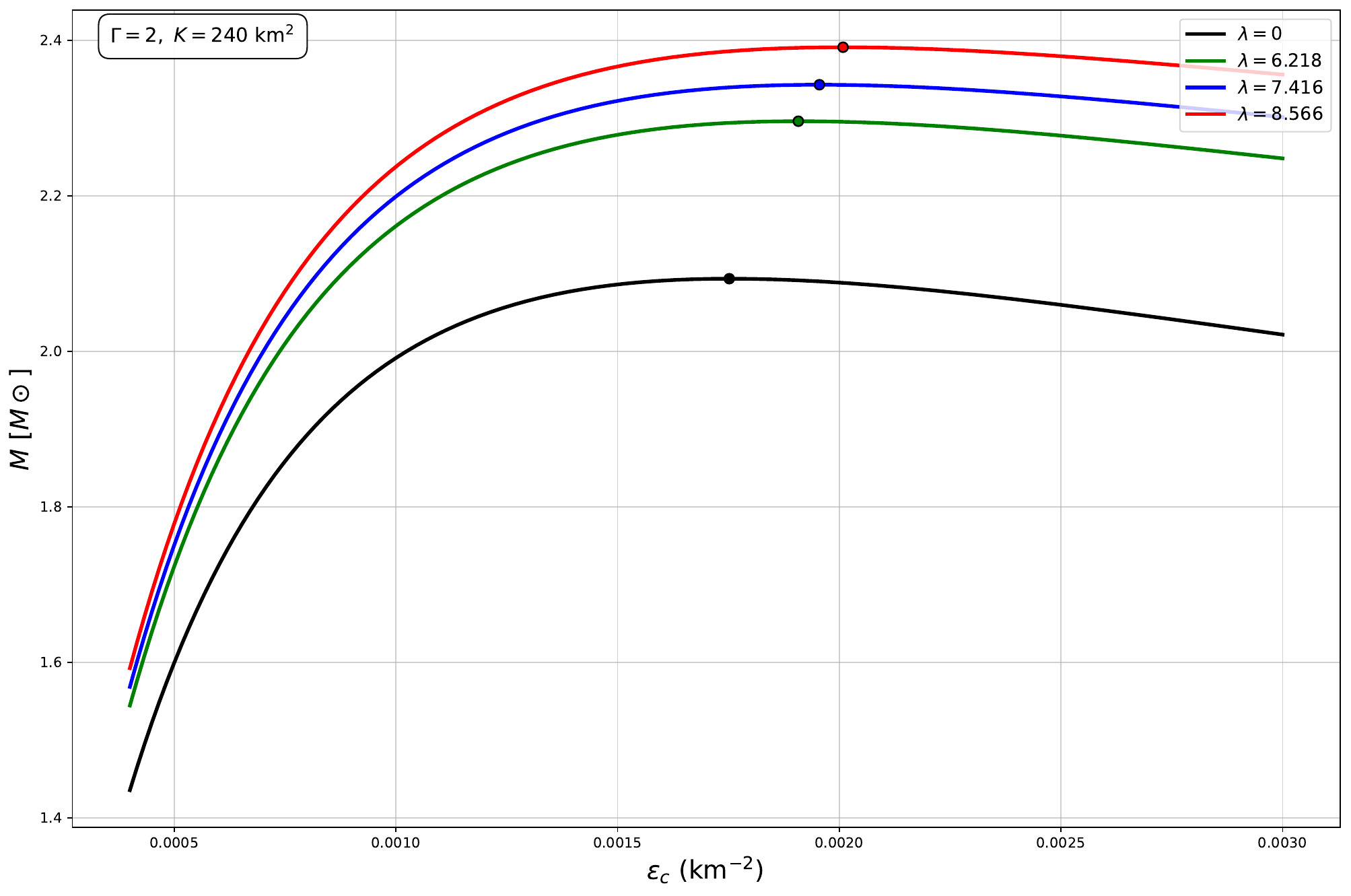}
\end{subfigure}

\end{figure*}

According to  Eq. \eqref{tovcc}, from \eqref{po11}--\eqref{po12} we obtain the modified TOV equation as (in geometric units $c=G=1$)
\begin{align}
&P'(r)= -\dfrac{m(r)\epsilon(r)}{ r^2}\left[1+\dfrac{P(r)}{\epsilon(r)} \right]\left[ 1-\dfrac{2m(r)}{ r}\right]^{-1}\times  \nonumber\\
& \left\lbrace 1+\dfrac{4\pi r^3P(r)}{m(r)}-\frac{ r^3\left[2\Lambda +\lambda K\epsilon^2(r)\left( 1+\dfrac{2}{3}K\epsilon(r)\right)  \right] }{4m(r)} \right\rbrace, \label{tovp}
\end{align}
with
\begin{align}
&m(r)= 4\pi\int_0^r r^2\epsilon(r)dr \nonumber\\
 &+\dfrac{1}{4}\int_0^r r^2\left[2\Lambda-\lambda\epsilon(r)\left( 1+K\epsilon(r)+\frac{1}{3}K^2\epsilon^2(r)\right)  \right] dr \label{mp}
\end{align}
and the equation of state 
\begin{align}
P(r)=K\epsilon^2(r). \label{pk}
\end{align}
Here, the two unknown constants are $K$ from the equation of state \eqref{pk} and $\lambda$ from the modified gravity contribution in $f(R,T)$. Following the numerical procedure described below Eq.~\eqref{tovc}, we obtain the results presented in Fig. \ref{fig:6in1} and  Table \ref{tab:3in1}. Where the values of the constants \(\lambda\) are selected such that this model alleviates the Hubble tension in cosmology.  When solving the system of Eqs. \eqref{tovp}-\eqref{mp}, it is easy to see that, in the context of studying matter inside stars, $\Lambda$ is extremely small and can be neglected (in geometric units $\Lambda \sim  10^{-46}~\mathrm{km}^{-2}$). It only becomes significant when describing the evolution of the Universe.\\

Observational constraints from NICER, gravitational-wave events, and precise pulsar mass measurements indicate that the maximum mass of neutron stars ($ M_{\rm max} $ or $M_{\rm TOV}$) lies in the range of approximately $2.2 - 2.35\,M_\odot$ \cite{Romani:2025ytn, Fan:2023spm, Annala:2017llu}.

In General Relativity, for the single polytropic equation of state $P = K \epsilon^2$, reproducing the observed maximum neutron star masses in the range $2.2 - 2.35\,M_\odot$ typically requires relatively large values of the constant $K$. However, such large values of $K$ lead to a violation of the trace energy condition (TEC) in GR. The trace energy condition (TEC) is
\begin{align}
T = \epsilon - 3P = \epsilon - 3K\epsilon^2 > 0, \label{dknl}
\end{align}
even though GR does not violate the causality condition on the speed of sound \cite{Ozel:2016oaf},
\begin{align}
v_s^2 = \frac{dP}{d\epsilon} = 2K\epsilon < 1. \label{dkat}
\end{align}
As clearly shown in Table \ref{tab:3in1}, the GR case ($\lambda=0$) indeed violates the energy condition \eqref{dknl} for the adopted values of $K$. The causality condition on the speed of sound is a strict requirement that must not be violated. However, there is no strong theoretical constraint demanding the satisfaction of the trace energy condition (TEC). For the chosen values of \(K\) and \(\lambda\), the present model increases the degree of violation of the TEC, but still fully respects the causality condition on the speed of sound.

Unlike the stellar mass $ M_{\rm max} $, which can be measured nearly directly, the determination of the stellar radius $ R_{M_{\rm max}} $ strongly depends on the underlying gravitational theory. For instance, within the framework of General Relativity, the radius of neutron stars $ R_{M_{\rm max}} $ is typically constrained to the range $12$--$13$ km from NICER observations \cite{Miller:2021qha, Dittmann:2024mbo} and constraints from the gravitational-wave event GW170817 \cite{LIGOScientific:2018cki, Raithel:2018ncd}.

We find that the cases with ($K=230$, $\Gamma=2$, $B=1$, $\lambda=7.416$) and ($K=240$, $\Gamma=2$, $B=1$, $\lambda=7.416$) provide  good agreement with current observational constraints on $ M_{\rm max} $ and $ R_{M_{\rm max}} $. It should be emphasized that this is only one illustrative example demonstrating that the present modified gravity model can fit the observational data. In fact, within the framework of the current model, one can choose different combinations of the parameters $K$, $L$, $B$, and $\lambda$ to obtain $M$--$R$ relations that better match future, more precise observational data.

In Fig.~\ref{fig:6in1} and Table~\ref{tab:3in1}, it can be seen that as \(\lambda\) increases, both the maximum mass \(M_{\rm max}\) and the corresponding radius \(R_{M_{\rm max}}\) increase, while the central energy density \(\epsilon_c\) also rises. This implies that the radius grows more slowly than the mass, as further demonstrated by the columns \(\Delta M_{\rm max}\) and \(\Delta R_{\rm Mmax}\) in Table~\ref{tab:3in1}, leading to more compact neutron star configurations.

\textbf{\subsection{Application to Cosmology}}

\subsubsection{Conservation Law \(\nabla_\mu T^\mu_{~\nu}=0\) in the Universe}

Previous studies in \(f(R,T)\) gravity, which employed an inaccurate expression for the tensor \(\Theta_{\mu\nu}\), concluded that only a restricted class of functions \(f(R,T)\) satisfies the conservation law \(\nabla_\mu T^\mu_{~\nu}=0\) (e.g., see \cite{Bertini:2023pmp, Alvarenga:2013syu, Baffou:2013dpa}). We now apply the more accurate expression for \(\Theta_{\mu\nu}\) obtained from Eq.~\eqref{T7} to cosmology to see what results emerge.\\[2mm]
We use the flat Friedmann-Lemaître-Robertson-Walker (FLRW) metric (setting \(c=1\))
\begin{align}
ds^2=dt^2-a^2(t)\left( dx^2 + dy^2 + dz^2 \right). \label{a39}
\end{align}
With the metric \eqref{a39},  the non-vanishing components of the Christoffel symbols are
\begin{align}
&\Gamma^0_{ij}=-\frac{\dot{a}}{a}g_{ij}, \label{ch1}\\
&\Gamma^i_{0j}=\frac{\dot{a}}{a}\delta^i_j, \label{ch2}
\end{align}
with $ i, j=(1, 2, 3) $. And the non-vanishing components of the Ricci tensor are
\begin{align}
R^0_{~0}&=3\frac{\ddot{a}}{a}, \label{a42}\\
R^i_{~j}&=\left( \frac{\ddot{a}}{a}+2\frac{\dot{a}^2}{a^2}\right)\delta^i_{~j}. \label{a43}
\end{align}
for $i,j=1,2,3$. Consequently, the Ricci scalar reads
\begin{align}
R=6\left( \frac{\ddot{a}}{a}+\frac{\dot{a}^2}{a^2}\right)=6\left(A+ H^2 \right), \label{a44}
\end{align}
where $H=\dfrac{\dot{a}}{a}$ is the Hubble parameter and we have introduced the notation $A=\dfrac{\ddot{a}}{a}$.

The components of the energy-momentum tensor in the comoving frame are
\begin{align}
&T^0_{~0}=\epsilon, \label{vtf1}\\
&T^1_{~1}=T^2_{~2}=T^3_{~3}=-P, \label{vtf2}
\end{align}
and
\begin{align}
\nabla_\mu T^\mu_{~0}=\dot{\epsilon}+3(\epsilon+P)\frac{\dot{a}}{a}. \label{dhT}
\end{align}

We define BCDM as the matter component with \(P=0\), i.e., its energy-momentum tensor \(T_{\mu\nu}\) takes the form of that of a perfect fluid with vanishing pressure. This includes baryonic matter and cold dark matter.

If the Universe contains only BCDM, we can set \(P=0\), which implies \(\epsilon = \rho\). Substituting Eqs.~\eqref{theta1}--\eqref{theta2} into Eq.~\eqref{rt9cu} then yields the modified continuity equation
\begin{align}
\nabla_\mu T^\mu_{~0} = \dot{\rho} + 3\rho \frac{\dot{a}}{a} 
= \frac{\frac{3}{2} \left( \dot{\rho} + 3\rho \frac{\dot{a}}{a} \right) f_T(R, T)}{k + f_T(R, T)}. \label{bt}
\end{align}
Here we have written the equation for the \(\nu=0\) component; the components \(\nu=1,2,3\) vanish automatically. Care must be taken when deriving Eq.~\eqref{bt}. For example, although \(\Theta^1_{~0}=0\), we have \(\nabla_1 \Theta^1_{~0} = (\Theta^0_{~0} - \Theta^1_{~1})\Gamma^1_{01}\). Equation \eqref{bt} can be rewritten as
\begin{align}
\nabla_\mu T^\mu_{~\nu} = \frac{\frac{3}{2} f_T(R, T) \nabla_\mu T^\mu_{~\nu}}{k + f_T(R, T)}. \label{btm}
\end{align}
This equation holds for \(\nu=0\). Since \(\nabla_\mu T^\mu_{~1} = \nabla_\mu T^\mu_{~2} = \nabla_\mu T^\mu_{~3} = 0\), it also holds for \(\nu=1,2,3\). The only solution of Eq.~\eqref{btm} is
\begin{align}
\nabla_\mu T^\mu_{~\nu} = 0. \label{btm2}
\end{align}
Thus, we conclude that for any function \(f(R,T)\), the theory always satisfies the conservation law \(\nabla_\mu T^\mu_{~\nu}=0\) during the matter-dominated era (when \(P=0\)).

We now prove that the conservation law \eqref{btm2} also holds during the radiation-dominated era. Using Eq.~\eqref{Tr} and noting that \(T=0\) for the electromagnetic field (see Eq.~\eqref{TEa}), Eq.~\eqref{rt9cu} becomes
\begin{align}
\nabla_\mu T^\mu_{~\nu} = \frac{f_T(R, T) \nabla_\mu T^\mu_{~\nu}}{k + f_T(R, T)}. \label{tr1}
\end{align}
This equation has the unique solution $\nabla_\mu T^\mu_{~\nu} = 0$. Hence, \(f(R,T)\) gravity also satisfies the conservation law during the radiation-dominated epoch.

If the Universe contains both BCDM and radiation, with their respective energy-momentum tensors \(T^{(m)}_{\mu\nu}\) and \(T^{(r)}_{\mu\nu}\), then \(T_{\mu\nu} = T^{(m)}_{\mu\nu} + T^{(r)}_{\mu\nu}\) and \(\Theta_{\mu\nu} = \Theta^{(m)}_{\mu\nu} + \Theta^{(r)}_{\mu\nu}\). Combining Eqs.~\eqref{theta1}--\eqref{theta2} for BCDM (with \(P=0\), so \(\epsilon=\rho\)) and Eq.~\eqref{Tr} for radiation, Eq.~\eqref{rt9cu} becomes
\begin{align}
\nabla_\mu \left( T^{(m)\mu}_{~\nu} + T^{(r)\mu}_{~\nu} \right) 
= \frac{f_T(R, T) \nabla_\mu \left( \frac{3}{2} T^{(m)\mu}_{~\nu} + T^{(r)\mu}_{~\nu} \right)}{k + f_T(R, T)}. \label{trm}
\end{align}
Assuming the radiation energy density is \(n\) times the BCDM density (\(n\) is a real number), i.e., \(T^{(r)\mu}_{~0} = n T^{(m)\mu}_{~0}\), we have
\[
\frac{3}{2} T^{(m)\mu}_{~0} + T^{(r)\mu}_{~0} = \frac{n + 3/2}{n+1} \left( T^{(m)\mu}_{~0} + T^{(r)\mu}_{~0} \right).
\]
Thus, Eq.~\eqref{trm} reduces to
\begin{align}
\nabla_\mu T^\mu_{~\nu} = \frac{\frac{n + 3/2}{n+1} f_T(R, T) \nabla_\mu T^\mu_{~\nu}}{k + f_T(R, T)}. \label{trm1}
\end{align}
This equation holds for \(\nu=0\), and since the spatial components vanish, it also holds for \(\nu=1,2,3\). From \eqref{trm1}, this equation has the unique solution
\begin{align}
\nabla_\mu T^\mu_{~\nu} = 0. \label{trm2}
\end{align}

It is also easy to see that, for the cosmological constant with \(T^\mu_{~\nu}=\dfrac{1}{k}\Lambda\delta^\mu_{~\nu}\), the conservation law \eqref{trm2} is likewise satisfied. Therefore, if the energy-momentum tensor \(T_{\mu\nu}\) of the Universe contains only standard matter components (baryonic matter, cold dark matter, radiation, and the cosmological constant as dark energy), then \(f(R,T)\) gravity satisfies the conservation law \(\nabla_\mu T^{\mu\nu} = 0\) for any function \(f(R,T)\). This is one of the main results of this paper.

It is also straightforward to verify that, when using the previous (incorrect) expressions for \(\Theta_{\mu\nu}\), the conservation law in the Universe holds only for a specific class of \(f(R,T)\) functions (e.g., see \cite{Bertini:2023pmp, Alvarenga:2013syu, Baffou:2013dpa}).

\subsubsection{Model \(f(R,T)\) Satisfying the Conservation Law \(\nabla_\mu T^\mu_{~\nu}=0\) for Both the Universe and Stars}

Using the metric \eqref{a39} and noting that

\(
\nabla^\mu\nabla_\nu f_R = g^{\mu\alpha}\nabla_\alpha\nabla_\nu f_R = g^{\mu\alpha}\left( \partial_\alpha\partial_\nu f_R - \Gamma^\beta_{\alpha\nu}\partial_\beta f_R \right),
\)
among the 16 equations contained in \eqref{rt6} only two are independent:
\begin{equation}
3\frac{\ddot{a}}{a}f_R-3\frac{\dot{a}}{a}\dot{f}_R-\frac{1}{2}f(R)=-k\epsilon-\left(\epsilon+\Theta^0_{~0} \right)f_T, \label{vtf3}
\end{equation}
\begin{equation}
\left(\frac{\ddot{a}}{a}+2\frac{\dot{a}^2}{a^2} \right)f_R -2\frac{\dot{a}}{a}\dot{f}_R-\ddot{f}_R-\frac{1}{2}f(R)=kP+\left(P-\Theta^1_{~1} \right)f_T, \label{vtf4}
\end{equation}
where \(\dot{f}_R \equiv \dfrac{\partial f_R(R, T)}{\partial t}\).

As shown above, if the energy-momentum tensor \(T_{\mu\nu}\) of the Universe contains only standard matter components (baryonic matter, cold dark matter, radiation, and the cosmological constant as dark energy), then the conservation law \(\nabla_\mu T^\mu_{~\nu}=0\) is satisfied for any function \(f(R,T)\). Therefore, it is sufficient for the function \(f(R,T)\) to satisfy the conservation law for astrophysical objects (such as stars) in order for it to also satisfy the conservation law in cosmology. We consider the simplest \(f(R,T)\) model presented in section \ref{star}, which takes the form \eqref{hn1}: \(f(R,T)=R+h(T)\) with
\begin{align}
h(T)=-2\Lambda+\lambda Q\left(1+KQ+\frac{1}{3}K^2Q^2 \right) \label{sc} 
\end{align}
and the relation between \(Q\) and \(T\) given by
\begin{align}
T=Q-3KQ^2. \label{sc1}
\end{align}
From Eqs.~\eqref{sc} and \eqref{sc1}, we readily obtain
\begin{align}
h_T(T)=\frac{\lambda\left(1+KQ\right)^2}{1-6KQ}. \label{sc2}
\end{align}
For Eq.~\eqref{sc1} to have real solutions, the following condition must hold:
\begin{align}
12KT \leq 1. \label{sc3}
\end{align}
Equation \eqref{sc1} has two solutions \(Q=\dfrac{1\pm \sqrt{1-12KT}}{6K}\). We choose the branch
\begin{align}
Q=\frac{1- \sqrt{1-12KT}}{6K}=\frac{2T}{1+\sqrt{1-12KT}}. \label{sc4}
\end{align}
This choice ensures that at late times (when \(T\) is small), \(Q \approx T\) (noting that \(T=\epsilon-3P\)). The other branch would yield \(Q \approx 1/(3K)\) at small \(T\), leading to a very large uncoupled constant \(\lambda/(3K)\) compared to \(\Lambda\), which is unsuitable for describing the Universe.

By dimensional analysis, with \([K]=[Q]^{-1}=[T]^{-1}=[\epsilon]^{-1}\) and \([\lambda Q]=[\lambda T]=[\lambda \epsilon]=[\Lambda]\), we have
\begin{align}
&K=10^6\frac{G}{c^4}K_{\rm Geo}, \label{ksi}\\
&\lambda=\frac{G}{c^4}\lambda_{\rm Geo}. \label{Lsi}
\end{align}
Here \(K\) and \(\lambda\) are in SI units, while \(K_{\rm Geo}\) and \(\lambda_{\rm Geo}\) are the corresponding values in geometric units (\(G=c=1\)) as shown in Table \ref{tab:3in1} of section \ref{star}.

If we consider only the epoch from matter domination to the present (neglecting the small contribution of radiation) and set \(P=0\), then combining Eqs.~\eqref{a44}, \eqref{theta1}--\eqref{theta2} and \eqref{sc}--\eqref{sc2}, the system of Eqs.~\eqref{vtf3}--\eqref{vtf4} is equivalent to the following system of equations (with \(\epsilon=\rho\) in units where \(c=1\))
\begin{align}
\frac{\dot{a}^2}{a^2}=\frac{k\rho}{3} +\frac{\Lambda}{3}-\frac{\lambda Q}{6}\left( 1+KQ+\frac{K^2Q^2}{3}\right), \label{sc5}
\end{align}
{\small
\begin{align}
\frac{\ddot{a}}{a}=-\frac{k\rho}{6} +\frac{\Lambda}{3}-\frac{\lambda Q}{6}\left( 1+KQ+\frac{K^2Q^2}{3}\right)+\frac{\lambda\rho(1+KQ)^2}{4(1-6KQ)}. \label{sc6}
\end{align}}
Retaining the term \(K\rho\) but neglecting \(K^2\rho^2\) compared to 1, and using Eq.~\eqref{sc4} (with \(P=0\) so \(T=\epsilon=\rho\)), we obtain
\begin{align}
Q=\rho(1+3K\rho). \label{sc7}
\end{align}
Substituting \eqref{sc7} and neglecting \(K^2Q^2\) relative to 1, Eqs.~\eqref{sc5}--\eqref{sc6} become (restoring \(c\))
\begin{align}
\frac{\dot{a}^2}{a^2}=\frac{1}{3}\left(k-\frac{\lambda}{2} \right)\rho c^4  +\frac{\Lambda c^2}{3}-\frac{2}{3}\lambda c^6 K\rho^2, \label{sc8}\\
\frac{\ddot{a}}{a}=-\frac{1}{6}\left(k-\frac{\lambda}{2} \right)\rho c^4  +\frac{\Lambda c^2}{3}+\frac{4}{3}\lambda c^6 K\rho^2. \label{sc9}
\end{align}
These can be rewritten as
\begin{align}
H^2=\frac{\dot{a}^2}{a^2}=\frac{\rho c^4}{3}k_{\rm eff}   +\frac{c^2\Lambda_{\rm eff}}{3}, \label{sc10}\\
A=\frac{\ddot{a}}{a}=-\frac{\rho c^4}{6}k_{\rm eff}   +\frac{c^2\Lambda_{\rm eff}}{3}. \label{sc11}
\end{align}
Where
\begin{align}
&k_{\rm eff}=k-\frac{\lambda}{2}-4\lambda c^2K\rho=k-\frac{\lambda}{2}-4\lambda c^2K\rho_0 (1+z)^3, \label{sc12}\\
&\Lambda_{\rm eff}=\Lambda+2\lambda c^4K\rho^2=\Lambda+2\lambda c^4K\rho^2_0(1+z)^6. \label{sc13}
\end{align}
In Eqs.~\eqref{sc12}--\eqref{sc13} we have used
\begin{align}
\rho=\left( \frac{a_0}{a}\right)^3 \rho_0= (1+z)^3\rho_0, \label{sc14}
\end{align}
which follows directly from the conservation law \(\nabla_\mu T^\mu_{~\nu}=0\) and Eq.~\eqref{dhT}. Here \(z = a_0/a - 1\) is the redshift and the subscript 0 denotes present-day values.

The equations \eqref{sc10}--\eqref{sc11} differ from the \(\Lambda\)CDM model in that the gravitational constant \(G\) and cosmological constant \(\Lambda\) are replaced by effective quantities. Since \(k=8\pi G/c^4\), Eq.~\eqref{sc12} gives the effective gravitational constant
\begin{align}
G_{\rm eff}=G - \frac{\lambda c^4}{16\pi}-\frac{\lambda c^6K\rho_0}{2\pi}(1+z)^3. \label{sc15}
\end{align}
Requiring \(G_{\rm eff}>0\) (attractive gravity) leads to
\begin{align}
\lambda < \frac{16\pi G}{c^4}. \label{sc16}
\end{align}
Comparing with \eqref{Lsi} yields
\begin{align}
\lambda_{\rm Geo} < 16\pi. \label{sc16a}
\end{align}
Noting the dimensional relations \eqref{ksi}, we observe that in Eqs.~\eqref{sc12}--\eqref{sc13} the last term can become comparable in magnitude to the second term only when \((1+z)^3 \sim \frac{c^2}{8\times 10^6 \rho_0 G K_{\rm Geo}}\). Combining this with the values of \(K_{\rm Geo}\) in Table \ref{tab:3in1} yields \(z \sim 10^{14}\). However, since we only consider redshifts in the range \(0 \lesssim z \lesssim 1100\), the last term can be safely neglected in Eqs.~\eqref{sc12}--\eqref{sc13}. Consequently, the effective parameters \(k_{\rm eff}\), \(\Lambda_{\rm eff}\), and \(G_{\rm eff}\) become constants:
\begin{align}
&k_{\rm eff}=k-\frac{\lambda}{2}, \label{sc17}\\
&\Lambda_{\rm eff}=\Lambda, \label{sc18}
\end{align}
and
\begin{align}
G_{\rm eff}=G - \frac{\lambda c^4}{16\pi}. \label{sc19}
\end{align}
Although \(k_{\rm eff}\), \(\Lambda_{\rm eff}\) and \(G_{\rm eff}\) are constants, \(k_{\rm eff}\) and \(\Lambda_{\rm eff}\) do not scale with the same factor. Therefore, this model is not a simple rescaling of \(\Lambda\)CDM and can produce a redshift-dependent quantity \(\mathbb{H}_0(z)\) (see the discussion below Eq.~\eqref{GRct} in the Introduction). Using Eqs.~\eqref{sc14}, \eqref{sc17} and \eqref{sc18}, we can rewrite Eqs.~\eqref{sc10}--\eqref{sc11} as
\begin{align}
&H(z)=\frac{\dot{a}}{a}=H_0\sqrt{(1+z)^3\Omega^{\rm eff}_{m, 0}+\Omega^{\rm eff}_{\Lambda, 0}}, \label{sc20}\\
&A(z)=\frac{\ddot{a}}{a}=H_0^2\left[-\frac{1}{2}(1+z)^3\Omega^{\rm eff}_{m, 0}+\Omega^{\rm eff}_{\Lambda, 0} \right]. \label{sc21}
\end{align}
Where \(H_0 = H(0)\) and
\begin{align}
&\Omega^{\rm eff}_{m, 0}=\frac{c^4\rho_0 k_{\rm eff}}{3H_0^2}, \label{sc22}\\
&\Omega^{\rm eff}_{\Lambda, 0} = \frac{c^2\Lambda}{3H_0^2}. \label{sc23}
\end{align}

To address the Hubble tension, modified gravity theories are expected to produce a redshift-dependent quantity \(\mathbb{H}_0(z)\) that decreases with increasing redshift, as discussed in the Introduction (see the text below Eq.~\eqref{GRct}). Following a simplified approach that can be found in Ref.~\cite{Krishnan:2020vaf}, we divide Eq.~\eqref{sc20} by the \(\Lambda\)CDM expansion factor $E^{\Lambda\rm CDM}(z) = \sqrt{(1+z)^3 \Omega^{\Lambda\rm CDM}_{m, 0} + \Omega^{\Lambda\rm CDM}_{\Lambda, 0}}$. This yields the following expression for \(\mathbb{H}_0(z)\) in the present model:
\begin{align}
\mathbb{H}_0(z) = H_0 \sqrt{\frac{(1+z)^3 \Omega^{\rm eff}_{m, 0} + \Omega^{\rm eff}_{\Lambda, 0}}{(1+z)^3 \Omega^{\Lambda\rm CDM}_{m, 0} + \Omega^{\Lambda\rm CDM}_{\Lambda, 0}}}. \label{sc24}
\end{align}
This equation describes the running of \(H_0\) with redshift \(z\), which is denoted as \(\mathbb{H}_0(z)\). Here
\begin{align}
&\Omega^{\Lambda\rm CDM}_{m, 0} = \frac{c^4 k \rho_0^{\Lambda\rm CDM}}{3\left(H_0^{\Lambda\rm CDM}\right)^2}, \label{sc25}\\
&\Omega^{\Lambda\rm CDM}_{\Lambda, 0} = \frac{c^2 \Lambda^{\Lambda\rm CDM}}{3\left(H_0^{\Lambda\rm CDM}\right)^2}. \label{sc26}
\end{align}
Neglecting the small contributions of radiation and neutrinos, we can take
\begin{align}
\Omega^{\Lambda\rm CDM}_{m, 0} + \Omega^{\Lambda\rm CDM}_{\Lambda, 0} = 1. \label{sc27}
\end{align}
To ensure \(\mathbb{H}_0(0) = H_0\), the same condition must hold for the effective parameters:
\begin{align}
\Omega^{\rm eff}_{m, 0} + \Omega^{\rm eff}_{\Lambda, 0} = 1. \label{sc28}
\end{align}
Using this condition, Eq.~\eqref{sc24} can be rewritten as
\begin{align}
\mathbb{H}_0(z) = H_0 \sqrt{\frac{(1+z)^3 \Omega^{\rm eff}_{m, 0} + 1 - \Omega^{\rm eff}_{m, 0}}{(1+z)^3 \Omega^{\Lambda\rm CDM}_{m, 0} + 1 - \Omega^{\Lambda\rm CDM}_{m, 0}}}. \label{sc29}
\end{align}
From Eq.~\eqref{sc29}, it is clear that \(H(0) = H_0 = \mathbb{H}_0(0)\). At large redshift, \(\mathbb{H}_0(z)\) approaches \(\mathbb{H}_0(z_{\rm large,\,CMB}) = H_0 \sqrt{\frac{\Omega^{\rm eff}_{m, 0}}{\Omega^{\Lambda\rm CDM}_{m, 0}}}\). 

Taking \(\Omega^{\Lambda\rm CDM}_{m, 0} = 0.315\) and \(\mathbb{H}_0(z_{\rm large,\,CMB}) = 67.4\)\\$\mathrm{km\,s^{-1}\,Mpc^{-1}}$ at \(z=1100\) (or large \(z\)) \cite{Planck:2018vyg}, we find that to obtain \(\mathbb{H}_0(z=0) = 73~\mathrm{km\,s^{-1}\,Mpc^{-1}}\) we must have
\begin{align}
&H_0 = 73~\mathrm{km\,s^{-1}\,Mpc^{-1}}, \label{sc30}\\
&\Omega^{\rm eff}_{m, 0} = 0.2685. \label{sc31}
\end{align}
Thus, \(H(0) = H_0 = \mathbb{H}_0(0) = 73~\mathrm{km\,s^{-1}\,Mpc^{-1}}\). It can also be easily shown that \(\dfrac{d\mathbb{H}_0(z)}{dz} < 0\), meaning \(\mathbb{H}_0(z)\) decreases as \(z\) increases. These results indicate that the model is capable of alleviating the Hubble tension.\\[2mm]
If we instead impose \(\mathbb{H}_0(0) = H_0^{\Lambda\rm CDM} = 73~\mathrm{km\,s^{-1}\,Mpc^{-1}}\) and \(\rho_0 = \rho_0^{\Lambda\rm CDM}\), then from Eqs.~\eqref{sc25} and \eqref{sc22} we obtain
\begin{align}
\frac{k_{\rm eff}}{k} = \frac{\Omega^{\rm eff}_{m, 0}}{\Omega^{\Lambda\rm CDM}_{m, 0}} = \frac{0.2685}{0.315}. \label{sc32}
\end{align}
Combining this with Eq.~\eqref{sc17} (noting that \(k = 8\pi G/c^4\)) gives
\begin{align}
\lambda = 7.416 \frac{G}{c^4}. \label{sc33}
\end{align}
Comparing with \eqref{Lsi} yields
\begin{align}
\lambda_{\rm Geo} = 7.416. \label{sc34}
\end{align}
These values satisfy the conditions \eqref{sc16}--\eqref{sc16a}.

\begin{figure*}[htbp]
\caption{\label{H0_eff_3_branches} \textit{The dependence of the redshift-dependent quantity \(\mathbb{H}_0(z)\) on redshift \(z\) is given by Eq.~\eqref{sc29}.}}
\begin{center}
\includegraphics[scale=0.5]{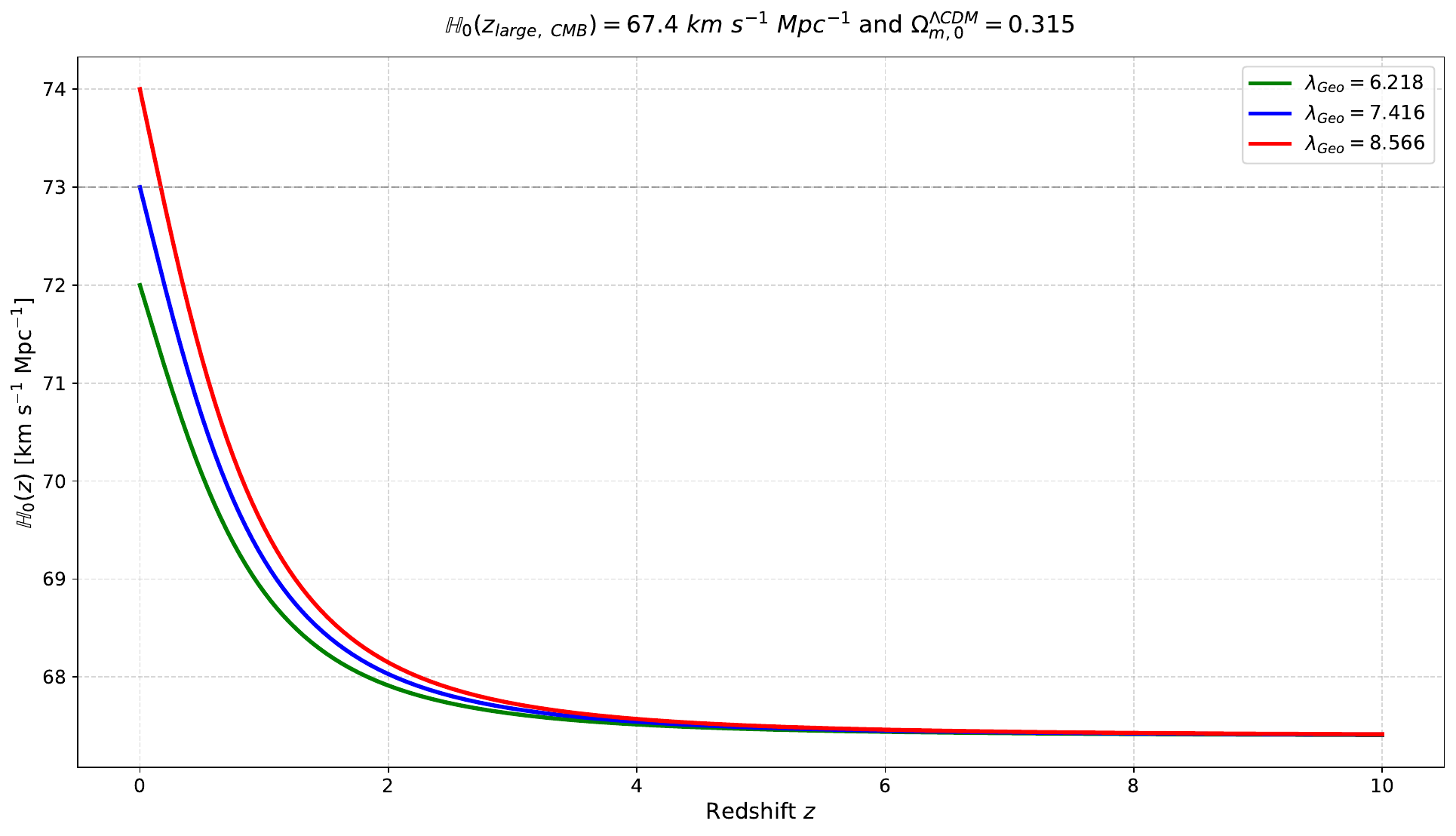}
\end{center}	
\end{figure*}

Performing the same calculation while keeping\\ \(\mathbb{H}_0(z_{\rm large,\,CMB}) = 67.4~\mathrm{km\,s^{-1}\,Mpc^{-1}}\) and \(\Omega^{\Lambda\rm CDM}_{m, 0} = 0.315\) fixed, we obtain the corresponding values of \(\Omega^{\rm eff}_{m,0}\) and \(\mathbb{H}_0(z=0)\) for each chosen \(\lambda_{\rm Geo}\) as shown in Table~\ref{tab:effective_parameters}.
\begin{table}[H]
\centering
\caption{\textit{Values of \(\Omega^{\rm eff}_{m,0}\) and \(\mathbb{H}_0(z=0)\) corresponding to the chosen values of \(\lambda_{\rm Geo}\). Note that \(H_0 = \mathbb{H}_0(z=0)\).}}
\label{tab:effective_parameters}
\begin{tabular}{c c c}
\toprule
$\lambda_{Geo}$ & ~~~~~$\Omega^{\rm eff}_{m,0}$ &~~~ $\mathbb{H}_0(z=0)$ \\
\midrule
6.218     & ~~~~0.2760                    & 72 \\
7.416     & ~~~~0.2685                    & 73 \\
8.566     & ~~~~0.2613                    & 74 \\
\bottomrule
\end{tabular}
\end{table}
The dependence of \(\mathbb{H}_0(z)\) on redshift \(z\) for the three values of \(\lambda_{Geo} = (6.218,~ 7.416,~ 8.566)\) is shown in Fig \ref{H0_eff_3_branches} and Table \ref{tab:Heff}, where Table \ref{tab:Heff} presents results only for the case \(\lambda_{\rm Geo} = 7.416\).

\begin{table*}[htbp]
\centering
\caption{\textit{Values of the redshift-dependent quantity \(\mathbb{H}_0(z)\) for the case \(\lambda_{\rm Geo} = 7.416\).}}
\label{tab:Heff}
\begin{tabular}{crcrcrcrcrcrcrcr}
\toprule
$z$ & $\mathbb{H}_0$ & ~~~$z$ & $\mathbb{H}_0$ & ~~~$z$ & $\mathbb{H}_0$ & ~~~$z$ & $\mathbb{H}_0$ & ~~~$z$ & $\mathbb{H}_0$ & ~~~$z$ & $\mathbb{H}_0$ & ~~~$z$ & $\mathbb{H}_0$ \\
\midrule
0.00 & $73.000$ & ~~~1.30 & $68.677$ & ~~~2.60 & $67.775$ & ~~~3.90 & $67.551$ & ~~~5.20 & $67.474$ & ~~~6.50 & $67.441$ & ~~~7.80 & $67.424$ \\
0.05 & $72.745$ & ~~~1.35 & $68.609$ & ~~~2.65 & $67.761$ & ~~~3.95 & $67.547$ & ~~~5.25 & $67.472$ & ~~~6.55 & $67.440$ & ~~~7.85 & $67.424$ \\
0.10 & $72.489$ & ~~~1.40 & $68.545$ & ~~~2.70 & $67.747$ & ~~~4.00 & $67.542$ & ~~~5.30 & $67.470$ & ~~~6.60 & $67.439$ & ~~~7.90 & $67.423$ \\
0.15 & $72.237$ & ~~~1.45 & $68.486$ & ~~~2.75 & $67.733$ & ~~~4.05 & $67.538$ & ~~~5.35 & $67.469$ & ~~~6.65 & $67.438$ & ~~~7.95 & $67.423$ \\
0.20 & $71.988$ & ~~~1.50 & $68.430$ & ~~~2.80 & $67.721$ & ~~~4.10 & $67.534$ & ~~~5.40 & $67.467$ & ~~~6.70 & $67.437$ & ~~~8.00 & $67.422$ \\
0.25 & $71.745$ & ~~~1.55 & $68.378$ & ~~~2.85 & $67.709$ & ~~~4.15 & $67.530$ & ~~~5.45 & $67.465$ & ~~~6.75 & $67.436$ & ~~~8.05 & $67.422$ \\
0.30 & $71.509$ & ~~~1.60 & $68.328$ & ~~~2.90 & $67.697$ & ~~~4.20 & $67.527$ & ~~~5.50 & $67.464$ & ~~~6.80 & $67.436$ & ~~~8.10 & $67.421$ \\
0.35 & $71.282$ & ~~~1.65 & $68.282$ & ~~~2.95 & $67.687$ & ~~~4.25 & $67.523$ & ~~~5.55 & $67.462$ & ~~~6.85 & $67.435$ & ~~~8.15 & $67.421$ \\
0.40 & $71.064$ & ~~~1.70 & $68.239$ & ~~~3.00 & $67.676$ & ~~~4.30 & $67.519$ & ~~~5.60 & $67.461$ & ~~~6.90 & $67.434$ & ~~~8.20 & $67.421$ \\
0.45 & $70.855$ & ~~~1.75 & $68.199$ & ~~~3.05 & $67.666$ & ~~~4.35 & $67.516$ & ~~~5.65 & $67.459$ & ~~~6.95 & $67.434$ & ~~~8.25 & $67.420$ \\
0.50 & $70.657$ & ~~~1.80 & $68.160$ & ~~~3.10 & $67.657$ & ~~~4.40 & $67.513$ & ~~~5.70 & $67.458$ & ~~~7.00 & $67.433$ & ~~~8.30 & $67.420$ \\
0.55 & $70.468$ & ~~~1.85 & $68.125$ & ~~~3.15 & $67.648$ & ~~~4.45 & $67.510$ & ~~~5.75 & $67.457$ & ~~~7.05 & $67.432$ & ~~~8.35 & $67.419$ \\
0.60 & $70.289$ & ~~~1.90 & $68.091$ & ~~~3.20 & $67.639$ & ~~~4.50 & $67.507$ & ~~~5.80 & $67.455$ & ~~~7.10 & $67.432$ & ~~~8.40 & $67.419$ \\
0.65 & $70.121$ & ~~~1.95 & $68.059$ & ~~~3.25 & $67.631$ & ~~~4.55 & $67.504$ & ~~~5.85 & $67.454$ & ~~~7.15 & $67.431$ & ~~~8.45 & $67.419$ \\
0.70 & $69.962$ & ~~~2.00 & $68.029$ & ~~~3.30 & $67.623$ & ~~~4.60 & $67.501$ & ~~~5.90 & $67.453$ & ~~~7.20 & $67.430$ & ~~~8.50 & $67.418$ \\
0.75 & $69.813$ & ~~~2.05 & $68.001$ & ~~~3.35 & $67.616$ & ~~~4.65 & $67.498$ & ~~~5.95 & $67.452$ & ~~~7.25 & $67.430$ & ~~~8.55 & $67.418$ \\
0.80 & $69.672$ & ~~~2.10 & $67.974$ & ~~~3.40 & $67.609$ & ~~~4.70 & $67.496$ & ~~~6.00 & $67.451$ & ~~~7.30 & $67.429$ & ~~~8.60 & $67.418$ \\
0.85 & $69.541$ & ~~~2.15 & $67.949$ & ~~~3.45 & $67.602$ & ~~~4.75 & $67.493$ & ~~~6.05 & $67.449$ & ~~~7.35 & $67.429$ & ~~~8.65 & $67.417$ \\
0.90 & $69.417$ & ~~~2.20 & $67.925$ & ~~~3.50 & $67.595$ & ~~~4.80 & $67.491$ & ~~~6.10 & $67.448$ & ~~~7.40 & $67.428$ & ~~~8.70 & $67.417$ \\
0.95 & $69.302$ & ~~~2.25 & $67.903$ & ~~~3.55 & $67.589$ & ~~~4.85 & $67.488$ & ~~~6.15 & $67.447$ & ~~~7.45 & $67.427$ & ~~~8.75 & $67.417$ \\
1.00 & $69.194$ & ~~~2.30 & $67.881$ & ~~~3.60 & $67.583$ & ~~~4.90 & $67.486$ & ~~~6.20 & $67.446$ & ~~~7.50 & $67.427$ & ~~~8.80 & $67.417$ \\
1.05 & $69.093$ & ~~~2.35 & $67.861$ & ~~~3.65 & $67.577$ & ~~~4.95 & $67.484$ & ~~~6.25 & $67.445$ & ~~~7.55 & $67.426$ & ~~~8.85 & $67.416$ \\
1.10 & $68.998$ & ~~~2.40 & $67.842$ & ~~~3.70 & $67.571$ & ~~~5.00 & $67.482$ & ~~~6.30 & $67.444$ & ~~~7.60 & $67.426$ & ~~~8.90 & $67.416$ \\
1.15 & $68.909$ & ~~~2.45 & $67.824$ & ~~~3.75 & $67.566$ & ~~~5.05 & $67.480$ & ~~~6.35 & $67.443$ & ~~~7.65 & $67.425$ & ~~~8.95 & $67.416$ \\
1.20 & $68.827$ & ~~~2.50 & $67.807$ & ~~~3.80 & $67.561$ & ~~~5.10 & $67.478$ & ~~~6.40 & $67.442$ & ~~~7.70 & $67.425$ & ~~~9.00 & $67.415$ \\
1.25 & $68.749$ & ~~~2.55 & $67.791$ & ~~~3.85 & $67.556$ & ~~~5.15 & $67.476$ & ~~~6.45 & $67.441$ & ~~~7.75 & $67.424$ & ~~~9.05 & $67.415$ \\
\bottomrule
\end{tabular}
\end{table*}

From Fig~\ref{H0_eff_3_branches} and Table~\ref{tab:Heff}, \(\mathbb{H}_0(z)\) in this model shows reasonable agreement with current observational data across a wide range of redshifts. At \(z \approx 0\), the model yields \(\mathbb{H}_0 = 73~\mathrm{km\,s^{-1}\,Mpc^{-1}}\), which is consistent with the latest local distance-ladder measurements ($73.0 - 73.5$) \cite{Riess:2021jrx}. At \(z \approx 0.5\), the value is $70.66$. At \(z=1\), this model predicts $69.19$, which lies slightly above but still within the broad range inferred from BAO \cite{Guo:2024pkx}. By \(z=2\), \(\mathbb{H}_0(z)\) decreases to $68.03$. From \(z=3\) ($\mathbb{H}_0=67.68$) onward, the parameter continues to decrease slowly and stabilizes, $67.48$ for \(z \approx 5\) and in good agreement with the CMB-inferred value of $ 67.4$ for $ z=1100 $ \cite{Planck:2018vyg}. 

 Overall, as shown in Fig~\ref{H0_eff_3_branches} and Table~\ref{tab:Heff}, the present model provides a consistent phenomenological description of the decreasing trend of \(\mathbb{H}_0(z)\) with increasing redshift. This behavior of \(\mathbb{H}_0(z)\) reflects the phenomenon of the running of \(H_0\) according to Eq.~\eqref{sc29}, in which the inferred value of \(H_0\) depends on the redshift \(z\) at which it is measured. In particular, \(\mathbb{H}_0(z)\) decreases rapidly at low redshifts (\(z \lesssim 1\)), more gradually in the intermediate range (\(1 \lesssim z \lesssim 2\)), and becomes nearly constant at high redshifts. This allows the model to alleviate the Hubble tension by bridging late- and early-Universe observations.\\

In the context of the Hubble tension, a distinctive aspect of this framework is that the relevant perturbative corrections are controlled by a single free parameter $\lambda$. This provides a relatively economical description with only one adjustable parameter in the late-time analysis, which may help reduce the risk of over-fitting compared to many multi-parameter phenomenological models \cite{DiValentino:2021izs, Vagnozzi:2023nrq, Abdalla:2022yfr, Dubey:2025khz}.\\[2mm]
A comparison with other modified gravity proposals aimed at resolving the Hubble tension--such as  $f(R)$ gravity \cite{Montani:2024xys}, teleparallel $f(T)$ gravity \cite{Verma:2025lmr}, simpler $f(R,T)$ models \cite{Dubey:2025khz, Aljohani:2025wsc}, and early dark energy (EDE) scenarios \cite{Takahashi:2022cpn}---shows several advantages of the present framework. Our model is significantly simpler, as it relies only on standard matter content without introducing additional scalar fields. It also avoids several common drawbacks of EDE models, such as shifts in the sound horizon and tensions with Big Bang nucleosynthesis (BBN) constraints. Moreover, as demonstrated in the previous sections, the same single parameter set simultaneously alleviates the Hubble tension at cosmological scales and reproduces the observed mass-radius (M--R) relation of neutron stars, providing a consistent description across both cosmological and high-density astrophysical regimes.\\

From the definitions \( H = \dfrac{\dot{a}}{a} = \dfrac{1}{a}\dfrac{da}{dt} \) and \( z = \dfrac{a_0}{a} - 1 \), it is straightforward to show that \( dt = -\dfrac{dz}{(1+z)H(z)} \). Therefore, the age of the Universe at a given redshift \(z\) is
\begin{align}
t = -\int_{+\infty}^{z} \dfrac{dx}{(1+x) H(x)}. \label{tuoi}
\end{align}
Here the function \(H(x)\) is computed from Eq.~\eqref{sc20} using the values of \(\Omega^{\rm eff}_{m,0}\) and \(H_0\) given in Table \ref{tab:effective_parameters} and taking into account Eq.~\eqref{sc28}. Note that \(1~{\rm Mpc} = 3.0856776 \times 10^{19}~{\rm km}\). The present age of the Universe \(t_{\rm today}\) is obtained by setting the upper limit of the integral to \(z=0\). To find the age at the onset of cosmic acceleration \(t_{\rm ac}\), we set \(A=0\) in Eq.~\eqref{sc21}, which gives
\begin{align}
z = \left( \frac{2 \Omega^{\rm eff}_{\Lambda, 0}}{\Omega^{\rm eff}_{m, 0}} \right)^{1/3} - 1. \label{tz}
\end{align}
The resulting ages of the Universe are presented in Table \ref{tab:tuoi}.

\begin{table}[H]
\centering
\caption{\textit{The present age of the Universe ($t_{\rm today}$) at $z = 0$, and the age of the Universe at the onset of accelerated expansion ($t_{\rm ac}$) for the three corresponding values of $z_{\rm ac}$ associated with the three values of $\lambda_{Geo}$.}}
\label{tab:tuoi}
\begin{tabular}{c c c c}
\toprule
$\lambda_{Geo}$  & ~~~~$t_{\rm today} ~(\rm Gyr)$ at $z = 0$    ~~~~~~&$t_{\rm ac}~(\rm Gyr)$ &$z_{\rm ac}$  \\
\midrule
6.218     & 13.400              & 7.006  &$0.738$ \\
7.416     & 13.317              & 6.875  &$0.760$\\
8.566     & 13.235              & 6.749  &$0.782$\\
\bottomrule
\end{tabular}
\end{table}

To facilitate comparison with the standard \(\Lambda\)CDM model, the corresponding ages in the \(\Lambda\)CDM cosmology are given in Table \ref{tab:tuoiGR}:
\begin{table}[H]
\centering
\caption{\textit{The present-day age of the Universe ($t_{\rm today}^{\Lambda\rm CDM}$) at $z = 0$ and the age of the Universe at the onset of acceleration ($t_{\rm ac}^{\Lambda\rm CDM}$) at $z_{\rm ac} = 0.632$ in the $\Lambda$CDM model.}}
\label{tab:tuoiGR}
\begin{tabular}{c c c}
\toprule
$H_0$ & ~~~~~$t_{\rm today}^{\Lambda CDM} ~(\rm Gyr)$ &~~~ $t_{\rm ac}^{\Lambda CDM}~(\rm Gyr)$ \\
\midrule
73       & ~~~~12.738                    & 7.104 \\
67.4     & ~~~~13.796                    & 7.695 \\
\bottomrule
\end{tabular}
\end{table}

In the preceding analysis, we have examined the behavior of the quantity \(\mathbb{H}_0(z)\) (Eq.~\eqref{sc29}), demonstrating that the present model is capable of alleviating the Hubble tension. We now turn our attention to the Hubble parameter \(H(z)\) given by Eq.~\eqref{sc20}. 

\begin{figure*}[htbp]
\caption{\label{fig Hz} \textit{The dependence of the Hubble parameter \(H(z)\) on redshift $z$ according to equation \eqref{sc20}. Where $\lambda_{Geo} = 0$ corresponds to General Relativity (the $\Lambda$CDM model).}}
\begin{center}
\includegraphics[scale=0.45]{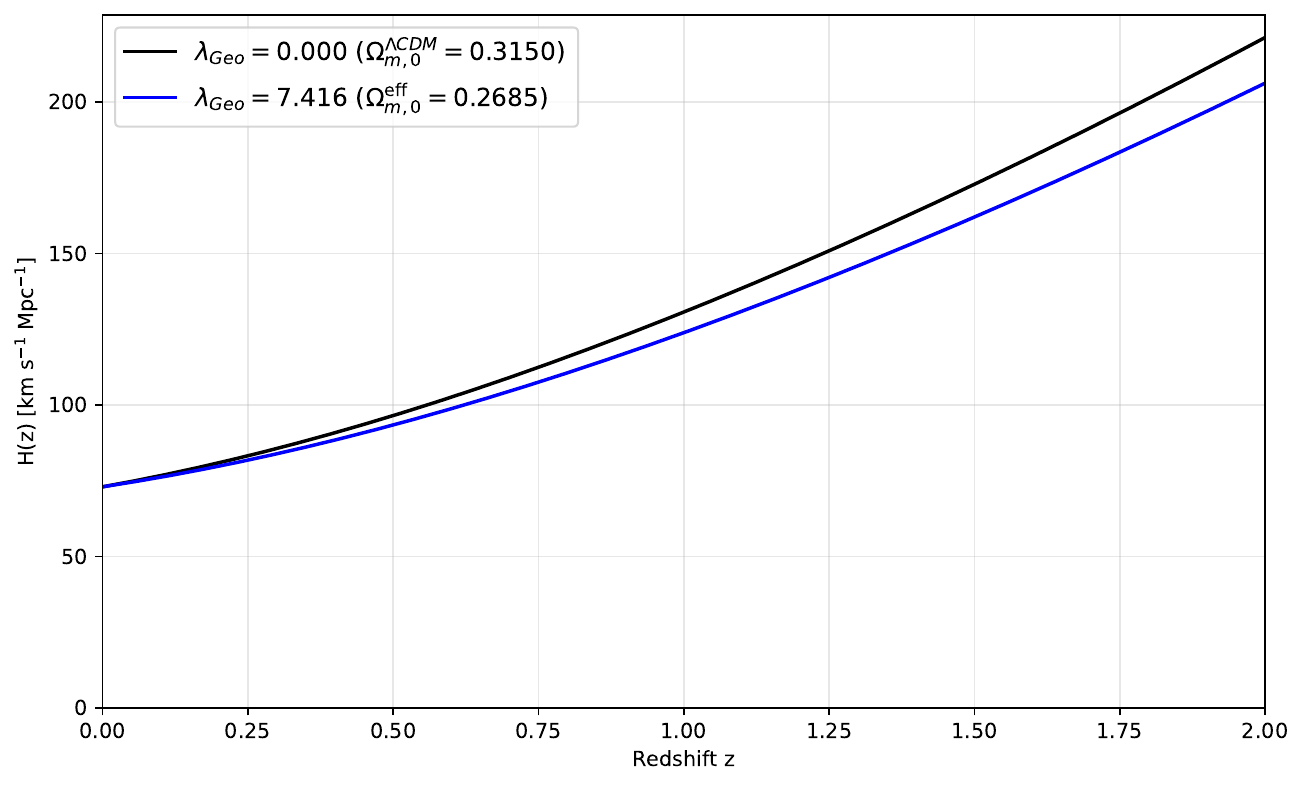}
\end{center}	
\end{figure*}

In contrast to \(\mathbb{H}_0(z)\), which decreases with increasing redshift \(z\), the Hubble parameter \(H(z)\) increases as \(z\) increases. Fig~\ref{fig Hz} and Table~\ref{tab Hz} show the redshift dependence of \(H(z)\) in the present modified gravity model with \(\lambda_{\rm Geo}=7.416\). As clearly seen in Fig.~\ref{fig Hz}, the difference between \(H(z)\) in our model and the corresponding value in the standard \(\Lambda\)CDM model becomes more pronounced at higher redshifts.\\

\begin{table*}[htbp]
\caption{\textit{The values of the Hubble parameter $H(z)$ for the model $\lambda_{\rm Geo}=7.416$ ($\Omega_{m,0}^{\rm eff}=0.2685$), and the values of the Hubble parameter $H^{\Lambda\rm CDM}(z)$ for $\lambda_{\rm Geo}=0$ ($\Omega_{m,0}^{\rm eff}=0.3150$) in General Relativity (the $\Lambda$CDM model). The last column in each group shows the relative difference in percent: $\Delta H\%=\left|H(z)-H^{\Lambda\rm CDM}(z)\right| / H^{\Lambda\rm CDM}(z) \times 100\%$.}}
\label{tab Hz}
\centering
\small
\begin{tabular}{rrrr|rrrr|rrrr}
\toprule
$z$ & $H(z)$ & $H^{\Lambda\rm CDM}(z)$ & $\Delta H\%$ & 
$z$ & $H(z)$ & $H^{\Lambda\rm CDM}(z)$ & $\Delta H\%$ &
$z$ & $H(z)$ & $H^{\Lambda\rm CDM}(z)$ & $\Delta H\%$ \\
\midrule
0.00 & 73.00 & 73.00 & 0.00\% &
0.70 & 104.54 & 109.08 & 4.16\% & 
1.40 & 153.88 & 163.88 & 6.10\% \\

0.05 & 74.53 & 74.79 & 0.35\% &
0.75 & 107.55 & 112.46 & 4.37\% &
1.45 & 157.92 & 168.33 & 6.18\% \\

0.10 & 76.17 & 76.71 & 0.70\% &
0.80 & 110.65 & 115.93 & 4.56\% & 
1.50 & 162.03 & 172.86 & 6.26\% \\

0.15 & 77.94 & 78.76 & 1.05\% &
0.85 & 113.83 & 119.49 & 4.74\% & 
1.55 & 166.20 & 177.44 & 6.33\% \\

0.20 & 79.82 & 80.94 & 1.39\% &
0.90 & 117.10 & 123.14 & 4.91\% &
1.60 & 170.43 & 182.08 & 6.40\% \\

0.25 & 81.81 & 83.24 & 1.72\% &
0.95 & 120.45 & 126.87 & 5.07\% &
1.65 & 174.72 & 186.79 & 6.46\% \\

0.30 & 83.91 & 85.66 & 2.04\% &
1.00 & 123.87 & 130.69 & 5.21\% &
1.70 & 179.06 & 191.55 & 6.52\% \\

0.35 & 86.13 & 88.21 & 2.35\% &
1.05 & 127.38 & 134.58 & 5.35\% &
1.75 & 183.45 & 196.37 & 6.58\% \\

0.40 & 88.46 & 90.87 & 2.65\% &
1.10 & 130.95 & 138.55 & 5.48\% &
1.80 & 187.90 & 201.25 & 6.63\% \\

0.45 & 90.89 & 93.64 & 2.94\% &
1.15 & 134.60 & 142.59 & 5.60\% &
1.85 & 192.41 & 206.18 & 6.68\% \\

0.50 & 93.42 & 96.52 & 3.21\% &
1.20 & 138.32 & 146.71 & 5.72\% &
1.90 & 196.96 & 211.16 & 6.72\% \\

0.55 & 96.05 & 99.51 & 3.47\% &
1.25 & 142.11 & 150.90 & 5.82\% &
1.95 & 201.57 & 216.21 & 6.77\% \\

0.60 & 98.79 & 102.60 & 3.71\% &
1.30 & 145.97 & 155.16 & 5.92\% &
2.00 & 206.23 & 221.30 & 6.81\% \\

0.65 & 101.62 & 105.79 & 3.94\% &
1.35 & 149.89 & 159.49 & 6.02\% & & & & \\
\bottomrule
\end{tabular}
\end{table*}

Table~\ref{tab Hz} allows a direct comparison with observational data. At \(z=0\), both the present model and \(\Lambda\)CDM give \(H(z)=H^{\Lambda\rm CDM}(z)=73~\mathrm{km\,s^{-1}\,Mpc^{-1}}\), consistent with local distance-ladder measurements \cite{Riess:2021jrx}. At \(z=0.43\), our model yields \(H(z)=89.9\), while \(\Lambda\)CDM gives \(H^{\Lambda\rm CDM}(z)=92.5\); both values agree with the model-independent measurement \(91.8 \pm 5.3\) reported in Ref.~\cite{Moresco:2016mzx}. At redshift \(z = 0.75\), Ref.~\cite{Jimenez:2023flo} reports a model-independent measurement of the Hubble parameter \(H(z) = 105.0 \pm 7.9\,(\rm stat) \pm 7.3\,(\rm sys)\); our model predicts \(H(z) = 107.55\), which agrees better with this observation than the \(\Lambda\)CDM value \(H^{\Lambda\rm CDM}(z) = 112.46\). At higher redshifts, for example at \(z=1.26\), the observed value of the Hubble parameter is \(135\pm 65\) according to Ref.~\cite{Tomasetti:2023kek}, and at \(z=1.965\) it is \(186.5\pm 50.4\) according to Ref.~\cite{Moresco:2015cya}. The uncertainties (\(\pm 65\) and \(\pm 50.4\)) are quite large; although the present modified gravity model is consistent with observations at these redshifts, more precise measurements with smaller uncertainties are still needed to better assess the accuracy of the model at high \(z\).\\

Additionally, we examine the deceleration parameter \(q(z)\), a dimensionless quantity defined as
\begin{align}
q(z)=-\frac{\ddot{a}a}{\dot{a}^2}= -\left(1+\frac{\dot{H}}{H^2} \right) =-\frac{A(z)}{H^2(z)}. \label{tsgt1}
\end{align}
where \(H(z)\) and \(A(z)\) are obtained from Eqs.~\eqref{sc20}--\eqref{sc21}.

\begin{figure*}[htbp]
\caption{\label{fig qz} \textit{The dependence of the  deceleration parameter \(q(z)\) on redshift $z$ according to equation \eqref{tsgt1}. Where $\lambda_{Geo} = 0$ corresponds to General Relativity (the $\Lambda$CDM model).}}
\begin{center}
\includegraphics[scale=0.45]{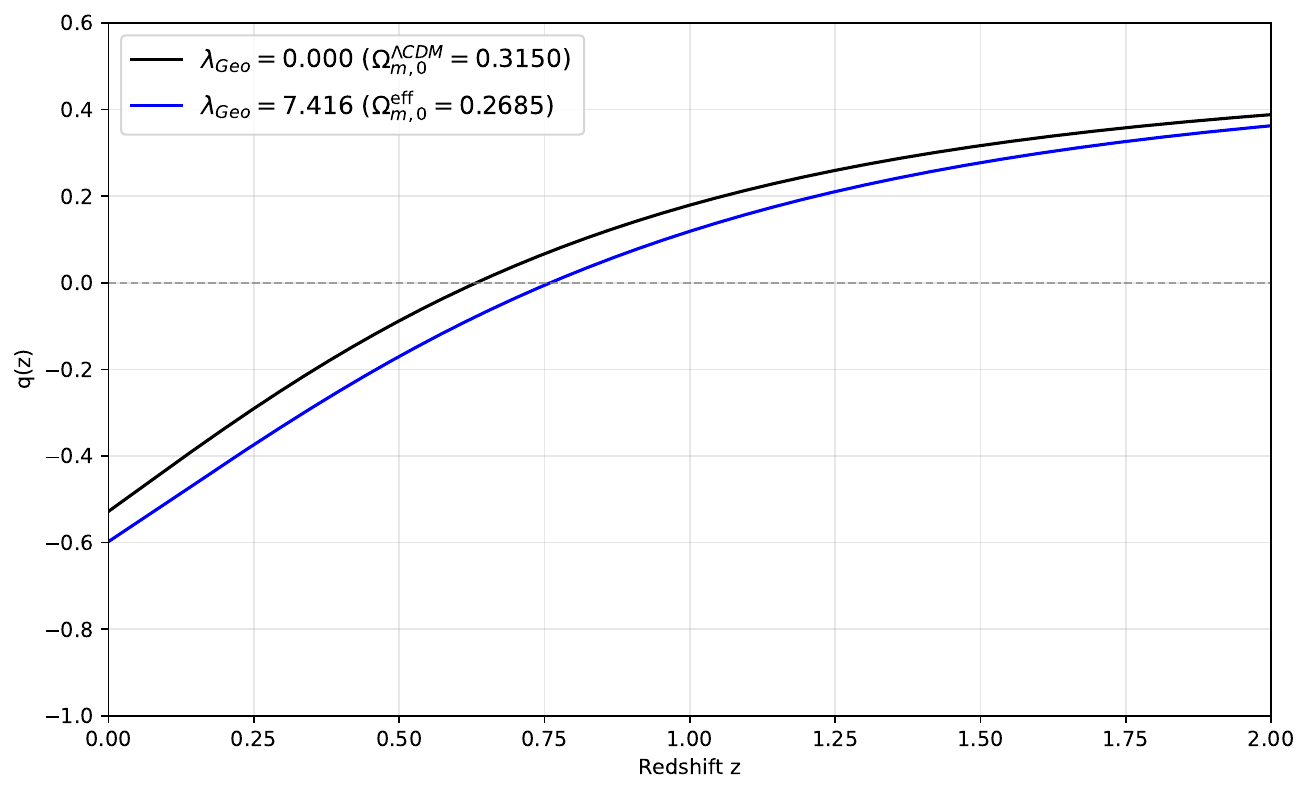}
\end{center}	
\end{figure*}

\begin{table*}[htbp]
\centering
\caption{\textit{The values of  the deceleration parameter $q(z)$ for the model $\lambda_{\rm Geo}=7.416$ ($\Omega_{m,0}^{\rm eff}=0.2685$), and the values of  the deceleration parameter $q^{\Lambda\rm CDM}(z)$ for $\lambda_{\rm Geo}=0$ ($\Omega_{m,0}^{\rm eff}=0.3150$) in General Relativity (the $\Lambda$CDM model). Where, $|\Delta q|=|q(z)-q^{\Lambda CDM}(z)|$ represents the absolute difference between $q(z)$ and $q^{\Lambda CDM}(z)$.}}
\label{tab qz}

\begin{tabular}{c c c c | c c c c}
\toprule
z & $q(z)$ & $q^{\Lambda CDM}(z)$ & $|\Delta q|$ 
& z & $q(z)$ & $q^{\Lambda CDM}(z)$ & $|\Delta q|$ \\
\midrule

0.0 & -0.5973 & -0.5275 & 0.0697 
& 1.1 & 0.1590 & 0.2148 & 0.0557 \\

0.1 & -0.5077 & -0.4305 & 0.0772 
& 1.2 & 0.1944 & 0.2456 & 0.0512 \\

0.2 & -0.4178 & -0.3358 & 0.0820 
& 1.3 & 0.2256 & 0.2726 & 0.0470 \\

0.3 & -0.3304 & -0.2462 & 0.0842 
& 1.4 & 0.2531 & 0.2961 & 0.0431 \\

0.4 & -0.2473 & -0.1632 & 0.0841 
& 1.5 & 0.2773 & 0.3167 & 0.0395 \\

0.5 & -0.1700 & -0.0878 & 0.0822 
& 1.6 & 0.2987 & 0.3348 & 0.0362 \\

0.6 & -0.0992 & -0.0202 & 0.0790 
& 1.7 & 0.3176 & 0.3508 & 0.0331 \\

0.7 & -0.0351 & 0.0398 & 0.0748 
& 1.8 & 0.3344 & 0.3648 & 0.0304 \\

0.8 & 0.0224 & 0.0926 & 0.0702 
& 1.9 & 0.3493 & 0.3772 & 0.0279 \\

0.9 & 0.0736 & 0.1389 & 0.0653 
& 2.0 & 0.3625 & 0.3882 & 0.0257 \\

1.0 & 0.1189 & 0.1794 & 0.0605 & & & & \\
\bottomrule
\end{tabular}
\end{table*}

Fig.~\ref{fig qz} and Table~\ref{tab qz} show the dependence of the deceleration parameter \(q(z)\) on redshift \(z\) according to Eq.~\eqref{tsgt1}. Note that time flows from right to left (corresponding to decreasing \(z\)). When \(q(z) > 0\), the universe is in a decelerating expansion phase (the expansion rate is slowing down). When \(q(z) < 0\), the universe is in an accelerating expansion phase (the expansion rate is speeding up). At low redshifts, as \(z\) increases, the difference in the deceleration parameter \(|\Delta q|\) between the present model and the \(\Lambda\)CDM model becomes larger. However, starting from \(z \approx 0.3\), this difference \(|\Delta q|\) decreases as \(z\) increases. For the present modified gravity model with \(\lambda_{\rm Geo}=7.416\), the universe transitions from deceleration to acceleration at \(z_{\rm ac} = 0.760\) (see Table~\ref{tab:tuoi} for the transition redshifts corresponding to other values of \(\lambda_{\rm Geo}\)). In the standard \(\Lambda\)CDM model, this transition occurs at \(z_{\rm ac} = 0.632\). Both values (\(z_{\rm ac}=0.760\) for our model and \(z_{\rm ac}=0.632\) for \(\Lambda\)CDM) lie within the observationally constrained range of approximately \(0.5\)--\(0.8\) reported in the model-independent analysis of Ref.~\cite{Myrzakulov:2024auu}. Similarly, both values are consistent with the average observed transition redshift \(0.64 \pm 0.16\) given in Ref.~\cite{Hu:2025fsz} (the transition redshift \(z_{\rm ac} = 0.64 \pm 0.16\) reported in Ref.~\cite{Hu:2025fsz} represents the average value). However, when using the Gaussian Process (GP) method, a model-independent reconstruction with \(H_0 \approx 73\), the transition occurs at \(z_{\rm ac} \approx 0.77\) (see Table 2 in Ref.~\cite{Hu:2025fsz}). This indicates that our modified gravity model is in better agreement with observations than the standard \(\Lambda\)CDM model.

\section{Conclusion}
In this work, we have derived a general expression for the metric variation \(\delta T_{\mu\nu}\) of the energy-momentum tensor of a perfect fluid under the assumption of particle number conservation. This derivation is performed directly from the standard form \(T_{\mu\nu} = (\epsilon + P) u_\mu u_\nu - P g_{\mu\nu}\), without imposing any restrictive conditions on the matter Lagrangian \(L_m\). The resulting expression, given in Eq.~\eqref{bpt}, is independent of the specific choice of \(L_m\) and can be readily applied to various modified theories of gravity.

Applying this result to \(f(R,T)\) gravity, we obtain a general expression for the tensor \(\Theta_{\mu\nu}\) [see Eq.~\eqref{T7}]. This expression is derived from the assumption of particle number conservation, but it has also been proven to be valid in the case of radiation (whether particle number is conserved or not). Using this \(\Theta_{\mu\nu}\), we have shown that in cosmology the conservation law \(\nabla_\mu T^{\mu\nu}=0\) is satisfied for  any function \(f(R,T)\).

Applying the obtained expression for \(\Theta_{\mu\nu}\) to the description of matter inside stars, we derive a class of functions that satisfy the conservation law, given in the general form \eqref{po6}. Considering the simplest case of the form \eqref{hn1}, we demonstrate that this model is capable of simultaneously alleviating the Hubble tension in cosmology and reproducing the mass-radius (\(M\)--\(R\)) relation consistent with current observational constraints on the maximum mass \(M_{\rm max}\) and the corresponding radius \(R_{M_{\rm max}}\) of neutron stars.\\

In addition to alleviating the Hubble tension, the present model also shows promising potential in addressing the $S_8$ tension. For instance, with $\lambda_{\rm Geo}=7.416$, Eqs.~\eqref{Lsi} and \eqref{sc19} yield an effective gravitational constant
\begin{equation}
\frac{G_{\rm eff}}{G} \approx 0.8525, \label{Geff}
\end{equation}
corresponding to a reduction of approximately $14.75\%$ compared to General Relativity (the $ \Lambda CDM $ model). The weaker gravity leads to slower growth of matter density perturbations, which can naturally suppress the value of $S_8$. The precise evaluation of \(\sigma_8\) and \(S_8\) in this model, which requires solving the linear perturbation equations, is reserved for future work. For a detailed introduction to the definitions and physical meaning of \(\sigma_8\) and \(S_8\), the reader is referred to the following references \cite{Abdalla:2022yfr, Hildebrandt:2016iqg}.\\

Future work will focus on extending this formalism to more realistic equations of state and exploring its implications for gravitational-wave astronomy and neutron star phenomenology.\\

In the present work, we assume that particle number conservation ($\nabla_\sigma (\rho u^\sigma) = 0$) remains valid and restrict our analysis to models satisfying $\nabla_\mu T^{\mu\nu} = 0$. This choice is made to ensure consistency and theoretical safety. In $f(R,T)$ gravity and other models with geometry-matter coupling, $\nabla_\mu T^{\mu\nu} \neq 0$ in general. As mentioned at the end of the Introduction, the equation \(\nabla_\mu T^{\mu\nu} = 0\) does not strictly represent the conservation of energy-momentum as its name suggests. The true conservation law of energy and momentum should be that of the total energy-momentum tensor, which includes both the matter energy-momentum tensor \(T^{\mu\nu}\) and the gravitational field energy-momentum tensor \(t^{\mu\nu}\). Whether $\nabla_\mu T^{\mu\nu} \neq 0$ necessarily leads to (or permits) a violation of particle number conservation requires more detailed investigation in future studies. If $\nabla_\mu T^{\mu\nu} \neq 0$ but particle number conservation $\nabla_\sigma (\rho u^\sigma) = 0$ can still be maintained (an assumption commonly adopted in many modified gravity theories with geometry-matter coupling, e.g., Refs.~\cite{Minazzoli:2012md, Harko:2010zi, Avelino:2024rub}),  then modified gravity models with $\nabla_\mu T^{\mu\nu} \neq 0$ remain valid and interesting candidates.

\section*{Acknowledgement}
This research is funded by the National Foundation for Science and Technology Development (NAFOSTED) of Vietnam under contract No. 103.01-2025.23.  We are grateful to the Referee for their careful reading of the manuscript and for the valuable suggestions and constructive comments, which have greatly helped to improve the clarity and quality of this paper.

\appendix

\section{Proof of \(\nabla_\mu T^\mu_{~\nu}=0\) \label{pla}}

Under an infinitesimal coordinate transformation \(x'^\mu = x^\mu + \xi^\mu(x^\gamma)\), where \(\xi^\mu\) is a small vector field, the inverse metric transforms as
\begin{align}
g'^{\mu\nu}(x'^\gamma) &= g^{\alpha\beta}(x^\gamma)\partial_\alpha x'^\mu \partial_\beta x'^\nu \nonumber\\
&= g^{\alpha\beta}(x^\gamma)\left(\delta^\mu_\alpha + \partial_\alpha\xi^\mu\right)\left(\delta^\nu_\beta + \partial_\beta\xi^\nu\right) \nonumber\\
&= g^{\mu\nu}(x^\gamma) + g^{\mu\beta}(x^\gamma)\partial_\beta\xi^\nu + g^{\nu\alpha}(x^\gamma)\partial_\alpha\xi^\mu. \label{g1}
\end{align}
Here, terms quadratic in \(\xi\) have been neglected. On the other hand, \(g'^{\mu\nu}(x'^\gamma) = g'^{\mu\nu}(x^\gamma + \xi^\gamma) = g'^{\mu\nu}(x^\gamma) + \xi^\sigma \partial_\sigma g'^{\mu\nu}(x^\gamma)\). Neglecting second-order terms, we have \(\xi^\sigma \partial_\sigma g'^{\mu\nu}(x^\gamma) = \xi^\sigma \partial_\sigma g^{\mu\nu}(x^\gamma)\). Thus,
\begin{align}
&g'^{\mu\nu}(x^\gamma) - g^{\mu\nu}(x^\gamma) =\nonumber\\
& -\xi^\sigma \partial_\sigma g^{\mu\nu}(x^\gamma) + g^{\mu\beta}(x^\gamma)\partial_\beta\xi^\nu + g^{\nu\alpha}(x^\gamma)\partial_\alpha\xi^\mu. \label{g2}
\end{align}
Since \(\delta g^{\mu\nu} = g'^{\mu\nu}(x^\gamma) - g^{\mu\nu}(x^\gamma)\), and it can be verified directly that \(\nabla^\mu\xi^\nu + \nabla^\nu\xi^\mu = -\xi^\sigma\partial_\sigma g^{\mu\nu} + g^{\mu\beta}\partial_\beta\xi^\nu + g^{\nu\alpha}\partial_\alpha\xi^\mu\) (where \(\nabla^\mu = g^{\mu\rho}\nabla_\rho\)), Eq.~\eqref{g2} can be rewritten as \cite{L2}
\begin{align}
\delta g^{\mu\nu} &= \nabla^\mu\xi^\nu + \nabla^\nu\xi^\mu, \label{g3}\\
\delta g_{\mu\nu} &= -\nabla_\mu\xi_\nu - \nabla_\nu\xi_\mu. \label{g4}
\end{align}

For the matter action \(S_m = \frac{1}{c}\int L_m \sqrt{-g}\,d^4x\), the variation is
\begin{align}
\delta S_m = \frac{1}{c}\int \delta(L_m \sqrt{-g})\,d^4x = \frac{1}{2c}\int T_{\mu\nu} \delta g^{\mu\nu} \sqrt{-g}\,dx^4, \label{g5}
\end{align}
where \(T_{\mu\nu} = \dfrac{2}{\sqrt{-g}} \dfrac{\delta(\sqrt{-g}L_m)}{\delta g^{\mu\nu}}\).

Substituting Eq.~\eqref{g3} into Eq.~\eqref{g5} and using the symmetry of \(T_{\mu\nu}\), we obtain
\begin{align}
\delta S_m &= \frac{1}{c}\int T_{\mu\nu} \nabla^\mu \xi^\nu \sqrt{-g}\,dx^4 \label{g6}\\
&= \frac{1}{c}\int T^\mu_{~\nu} \nabla_\mu \xi^\nu \sqrt{-g}\,dx^4 \label{g7}\\
&= \frac{1}{c}\int \nabla_\mu (T^\mu_{~\nu} \xi^\nu) \sqrt{-g}\,dx^4 - \frac{1}{c}\int \xi^\nu \nabla_\mu T^\mu_{~\nu} \sqrt{-g}\,dx^4. \label{g8}
\end{align}

The first integral on the right-hand side of Eq.~\eqref{g8} becomes a surface term:
\begin{align}
\int \nabla_\mu (T^\mu_{~\nu} \xi^\nu) \sqrt{-g}\,dx^4 = \int \frac{\partial}{\partial x^\mu} \left( T^\mu_{~\nu} \xi^\nu \sqrt{-g} \right) dx^4 \label{g9}
\end{align}
and vanishes upon integration by parts (assuming appropriate boundary conditions). Therefore, Eq.~\eqref{g8} reduces to \cite{L2}
\begin{align}
\delta S_m = -\frac{1}{c} \int \xi^\nu \nabla_\mu T^\mu_{~\nu} \sqrt{-g}\,dx^4. \label{g9}
\end{align}

If the matter field moves in a curved spacetime and its trajectory obeys the principle of least action, then \(\delta S_m = 0\). Since \(\xi^\nu\) is an arbitrary infinitesimal vector field, it follows from Eq.~\eqref{g9} that \cite{L2}
\begin{align}
\nabla_\mu T^\mu_{~\nu} = 0. \label{g10}
\end{align}
This conservation law holds for any type of matter and for any modified theory of gravity. It is valid both for the motion of astrophysical objects and for the universe as a whole.

It should be noted that the definition\\ \(T_{\mu\nu} = \dfrac{2}{\sqrt{-g}} \dfrac{\delta(\sqrt{-g} L_m)}{\delta g^{\mu\nu}}\) includes variations with respect to both the metric tensor and the matter fields (including derivatives of the metric and matter fields with respect to coordinates). However, the variation with respect to the matter fields is governed by the equations of motion of the matter in curved spacetime. Consequently, this contribution vanishes on-shell, leaving us with the expression in Eq.~\eqref{rt7}.

\end{document}